\documentclass{aa}                                   
\usepackage{graphicx}
\usepackage{txfonts}
\usepackage{natbib,twoopt}
\usepackage{float}
\usepackage{multirow}
\usepackage{mathptmx}

\DeclareSymbolFont{lettersA}{U}{txmia}{m}{it}
\SetSymbolFont{lettersA}{bold}{U}{txmia}{bx}{it}
\DeclareFontSubstitution{U}{txmia}{m}{it}
\DeclareMathSymbol{v}{\mathalpha}{lettersA}{"33}

\begin{document}
\title{Astrometric mock observations for determining the local dark matter density}


\author{Shigeki Inoue\inst{\ref{inst1},\ref{inst2}}\thanks{\email{shigeki.inoue@nao.ac.jp}} \and Naoteru Gouda\inst{\ref{inst1}}}

\institute{National Astronomical Observatory of Japan, Mitaka, Tokyo 181-8588, Japan\label{inst1} \and Korea Astronomy and Space Science Institute 776, Daedeokdae-ro, Yuseong-gu, Daejeon, 305-348, Republic of Korea\label{inst2}}

\date{Received February 15, 2013; accepted ???? ??, ????}

\abstract
    {To determine the local dark matter density of the solar system is a classical problem in astronomy. Recently, a novel method of determining the local dark matter density from stellar distribution and vertical velocity dispersion profiles perpendicular to the Galactic plane was devised. This method has the advantage of abolishing conventional approximations and using only a few assumptions.}
    {Our aims are to carefully scrutinize this method and to examine influences by uncertainties of astrometric observations. We discuss how the determinations of the local dark matter density vary with observational precisions on parallax, proper motion, and line-of-sight velocity measurements.}
    {To examine the influences by the observational imprecision, we created mock observation data for stars that are dynamical tracers based on an analytical galaxy model and applied parametrized observational errors to the mock data. We evaluated the accuracy of determining the dark matter density by applying the method to the mock data. In addition, we estimated a sample size and observational precision required to determine the dark matter density with accuracy.}
    {We find that the method is capable of determining the local dark matter density with accuracy if the sample size and observational precisions are satisfactory. The required sample size is approximately 6,000 stars. The random errors of parallaxes and proper motions can cause systematic overestimation of the dark matter density. We estimate the required precisions of the parallax measurements to be approximately $0.1$--$0.3$ milliarcseconds at $1~{\rm kpc}$ away from the Sun; the proper motion precisions do not seem to be as important as the parallaxes. Moreover, we find that the line-of-sight velocity errors can cause either underestimation or overestimation of the dark matter density, which is contingent on distance-dependence of the errors.}
    {From these results, we expect that using the \textit{Hipparcos} catalog would overestimate the local dark matter density because of the imprecise parallax measurements if this method is applied; however, we emphasize the capability of the method. We expect that \textit{Gaia} will provide data precise enough to determine the local dark matter density.}

    \maketitle
%

\section{Introduction}
\label{Intro}
In recent years, dedicated searches for candidates of dark matter (DM) particles have intensified. Experiments aiming at direct detection of the DM particles look for signals from recoil of DM particles with nuclei inside the detector \citep[e.g.,][and references therein]{s:07,b:12}. Thus, the event rates of the direct detection are clearly proportional to the local DM density (LDMD) around the solar system, $\rho_{\rm dm}^\odot$. This is why the problem of determining the LDMD at the solar position has recently been attracting a great deal of attention.

Because of the ``dark" nature, however, it is difficult to understand details of the structures of the DM halo. Usually, we have to resort to dynamical analyses of observable phenomena, such as motions of stars, gas, and satellite galaxies. For example, a rotation velocity curve of the Galactic disk is often used to measure the DM halo mass and density profile \citep[e.g.,][]{am:86,as:91,s:09,iwt:13}; \citet{s:12}\footnote{\citet{s:12} used an extrapolation of a Navarro-Frenk-White density profile, which is determined by fitting to outer Galactic regions. However, he mentioned that his rotation curve inside $R\sim20~{\rm kpc}$ cannot constrain the DM density profile.} has recently estimated that $\rho_{\rm dm}^\odot=0.00612\pm0.00080~{\rm M_\odot~pc^{-3}}$; the other recent works also proposed  $\rho_{\rm dm}^\odot\simeq0.01~{\rm M_\odot~pc^{-3}}$.\footnote{$1~{\rm M_\odot~pc^{-3}}\simeq38.0000~{\rm GeV~cm^{-3}}.$} However, the analyses using rotation curves posit global mass-modelings and a spherical assumption of the Galactic halo. If the Galactic DM halo is far from a round shape, the rotation curve analyses can lead to erroneous determinations since estimated LDMDs are spherically averaged densities \citep[e.g.,][]{zdk:09}. Cosmological simulations have indeed shown that DM halos are generally aspherical \citep[e.g.,][]{dc:91,afp:06,bef:07,kdm:07}. Moreover, recent studies on dynamical analyses of the Sagittarius stream have preferred a nearly oblate halo in which the minor axis is approximately aligned with the line of sight to the Galactic center (e.g., \citealt{lm:10,ilm:12,dw:13},\footnote{Some studies are, however, inconsistent with the result of the oblate halo, preferring an approximately spherical halo \citep[e.g.,][]{ili:01,fbe:06} and a prolate halo \citep[e.g.,][]{h:04A}.} although see \citealt{om:01,drv:13}). Thus, more accurate determination of the LDMD requires following a more direct methodology than modeling the whole halo.

Some studies have tried to determine the LDMD with analyses of local kinematics. For example, \citet{sng:10} used an equation of centrifugal equilibrium and an observed local radial gradient of the rotation curve: They determined $\rho_{\rm dm}^\odot=0.0113\pm0.0055~{\rm M_\odot~pc^{-3}}$. In addition, other methods of fitting various dynamical parameters, such as Galactic rotations, halo masses and thicknesses of gas layers also derived LDMDs of $\rho_{\rm dm}^\odot\simeq0.01~{\rm M_\odot~pc^{-3}}$ \citep{om:00,om:01,wb:10,cu:10}. These methods, relying on the Galactic rotations and its gradients, however, still posit axisymmetric assumptions.

As another approach, density distribution and vertical motions of disk stars near the Sun are frequently used to determine the LDMD \citep[e.g.,][]{o:32,b:84a,kg:89I}. This method is advantageous in terms of being independent from the global shape of the DM halo. Previous studies have determined that $\rho_{\rm dm}^\odot\hspace{0.3em}\raisebox{0.4ex}{$<$}\hspace{-0.75em}\raisebox{-.7ex}{$\sim$}\hspace{0.3em}0.01~{\rm M_\odot~pc^{-3}}$ \citep[e.g.,][]{brc:87,crb:89,kg:89II,kg:89III,k:91,ff:94,p:97,ccb:98,hf:04,bt:12,swe:12,zrv:12}\footnote{\citet{mcm:12} proposed an absence of DM around the Sun from their analysis of thick-disk stars. However, \citet{bt:12} claimed that this result significantly underestimates the LDMD because of a false assumption.} although some earlier works indicated a preference for much higher LDMDs \citep[e.g.,][]{o:32,o:60,h:60,b:84c,b:84b}.

Recently, \citet[][hereafter G11 and G12]{grl:11,glr:12} have devised a new method of determining the LDMD from stellar density profile and motions perpendicular to the Galactic disk. Their method was named the minimal assumption (MA) method and uses, as the name suggests, only a few assumptions and a flexible 15-component galaxy model (see Sect. \ref{MA}). With the MA method, they determined the LDMDs of $\rho_{\rm dm}^\odot=0.033^{+0.008}_{-0.009}~{\rm M_\odot~pc^{-3}}$ in G11 and $\rho_{\rm dm}^\odot=0.022^{+0.013}_{-0.015}~{\rm M_\odot~pc^{-3}}$ in G12. They seemed to show that the methods of \citet{hf:04} and \citet{kg:89II,kg:89III,kg:89I} have systematic biases and underestimate the LDMD due to unsuitable assumptions of separable distribution function and isothermality for tracer stars, whereas the MA method does not need these assumptions. They applied the MA method to $N$-body simulation data, and demonstrated that their method can determine the LDMD within the 90 \% confidence level. As our first aim, we examined the capability of the MA method by applying it to mock observation data generated with an analytical galaxy model.

Fortunately, launches of next-generation astrometry satellites are now approaching: \textit{Gaia} \citep{Perryman:01}, \textit{Nano-JASMINE} \citep[Japan Astrometry Satellite Mission for INfrared Exploration,][]{gouda:11}, and \textit{JMAPS} \citep[Joint Milli-Arcsecond Pathfinder Survey,][]{gdh:09}. These satellites will provide astrometric information of Galactic stars with unprecedented precision. Therefore, it is worth performing a feasibility study of determining the LDMD with astrometric data provided by them. We study the dependence of LDMD determinations on a sampling region and observational errors such as parallaxes, proper motions, and line-of-sight velocities (LOSVs). Our second aim is to evaluate a sample size and observational precisions required to determine the LDMD with accuracy.

The basic concept of the MA method is described in Sect. \ref{MA}. Our galaxy model for the mock observations and how to create the mock data are explained in Sect. \ref{model} and \ref{data}. A Markov chain Monte Carlo (MCMC) technique combined with the MA method is explained in Sect. \ref{MCMC}. Our results are shown in Sect. \ref{results}. Discussion and our conclusions are presented in Sect. \ref{discussion} and Sect. \ref{conclusions}.

\section{The MA method}
\label{MA}

\subsection{Basic equations and deduced galaxy models for the MA method}
\label{MA_model}
We assume that the Galaxy can be modeled as a superposition of multiple visible components, such as stellar and gaseous disks, stellar halo, and a dark component, i.e., $\rho_{\rm tot}\equiv\sum\rho_i+\rho_{\rm dm}$, where $\rho_i$ represents the \textit{mass} density of the $i$-th visible component, and $\rho_{\rm tot}$ and $\rho_{\rm dm}$ are the total and DM mass density. The MA method introduces only three assumptions:
\renewcommand{\labelenumi}{(\theenumi)}
\begin{enumerate}
\item the system is in equilibrium,
\item a ``tilt" term of the Jeans equation is negligible,
\item the DM density is constant in the region we consider.
\end{enumerate}
G11 have confirmed that all of these assumptions hold in an $N$-body simulation of a barred spiral galaxy.

From assumption (1), the $i$-th component satisfies the Jeans equation. Considering the direction perpendicular to the Galactic disk ($z$-axis) and ignoring the tilt term of assumption (2), the Jeans equation can be reduced to
\begin{equation}
\label{notilt_Jeans}
\sigma_{z,i}^2\frac{\partial\nu_i}{\partial z} + \nu_i\left(\frac{\partial\sigma_{z,i}^2}{\partial z} + \frac{\partial\Phi}{\partial z}\right) = 0,
\end{equation}
where $\nu_i$ is density of the $i$-th component under the total gravitational potential $\Phi$ and $\sigma_z$ is the vertical velocity dispersion. Integrating this equation, one can obtain a density profile of the $i$-th population as 
\begin{equation}
\label{Jeans_integ}
\frac{\nu_i(z)}{\nu_{i}(0)} = \frac{\sigma_{z,i}^2(0)}{\sigma_{z,i}^2(z)}\exp\left(-\int^z_0\frac{1}{\sigma_{z,i}^2(z')}\frac{\mathrm{d}\Phi}{\mathrm{d}z'}\mathrm{d}z'\right).
\end{equation}
$\sigma_{z,i}(0)$ are given as parameters, whereas the profiles of $\sigma_{z,i}(z)$ must be assumed to solve this equation. G11 introduced parametrized runs of $\sigma_{z,i}(z)$, and G12 assumed isothermality for the model. Following G12, we also assume that the model components are isothermal; however, we discuss the influence of different $\sigma_{z,i}(z)$ profiles on the calculation in Appendix \ref{appendix1}. With the isothermal assumption, Eq. (\ref{Jeans_integ}) becomes
\begin{equation}
\label{single_density}
\rho_i(z) = \rho_i(0)\frac{\sigma_{z,i}^2(0)}{\sigma_{z,i}^2(z)}\exp\left(-\frac{\Phi(z)}{\sigma_{z,i}^2(0)}\right).
\end{equation}
Here, $\rho_i$ is mass density: $\rho_i=\nu_im_i$, where $m_i$ is the mass-to-light ratio of the $i$-th population. $\rho_i(0)$ are given as parameters. Finally, one can describe the profile of the total observable density as $\rho_{\rm s}(z) = \sum_i\rho_i(z)$. Then, the total matter density is $\rho_{\rm tot}(z)=\rho_{\rm s}(z)+\rho_{\rm dm}$ from assumption (3). In addition, the total surface density of the observable matters is derived as $\Sigma_{\rm s}(z) = 2\int^z_0\rho_{\rm s}(z')dz'$.

Next, $\rho_{\rm tot}$ is connected to the total gravitational potential $\Phi$ via the Poisson equation. By ignoring the radial gradient of the Galactic rotation curve,\footnote{\citet{bab:12IV} estimated that the contribution of the gradient of the rotation curve to the local density is only $0.0002^{+0.0002}_{\--0.0025}~{\rm M_\odot~pc^{-3}}$.} this is transformed to 
\begin{equation}
\label{Poisson3}
\frac{\partial^2\Phi}{\partial z^2} = 4\pi G\left(\rho_{\rm s}(z)+\rho_{\rm dm}\right).
\end{equation}

One can now compute the potential of the model galaxy $\Phi$ by numerical integration of Eqs. (\ref{single_density}) and (\ref{Poisson3}) with a given parameter set of $\rho_{i}(0)$, $\sigma_{z,i}(0)$ and $\rho_{\rm dm}$ and boundary conditions of $\Phi=\frac{\partial \Phi}{\partial z}=0$ at $z=0$. The total number of the parameters is, therefore, $2n+1$, where $n$ is the number of visible components in the galaxy model. In this paper, we refer to the computed potential $\Phi$ as the ``deduced model".

\subsection{A tracer population}
\label{MA_tracer}
The reliability of a trial parameter set used in the above computation of $\Phi$ must be evaluated by comparing it with observations. To this end, the MA method prepares an observed sample of ``tracer" population stars. The tracer stars must share the same kinematic state. The preferred tracer stars are considered to be old (well-mixed), bright (observationally precise and accurate) and large in number (statistically reliable). 

We now assume that a sample of tracer stars is available, and we obtain density and velocity dispersion profiles $\nu_{\rm trac}^{\rm obs}(z)$ and $\sigma_{z, \rm trac}^{\rm obs}(z)$ in a range from $z_{\rm min}$ to $z_{\rm max}$. Substituting $\sigma_{z, \rm trac}^{\rm obs}(z)$ for $\sigma_{z,i}$ in Eq. (\ref{Jeans_integ}), one can \textit{deduce} a fall-off of the tracer density profile under the modeled potential $\Phi$ as
\begin{equation}
\label{ded_tracer}
\frac{\nu_{\rm trac}^{\rm ded}(z)}{\nu_{\rm trac}^{\rm ded}(z_{\rm min})} = \frac{\sigma_{z, \rm trac}^{2~\rm obs}(z_{\rm min})}{\sigma_{z, \rm trac}^{2~\rm obs}(z)}\exp\left(-\int^z_{z_{\rm min}}\frac{1}{\sigma_{z, \rm trac}^{2~\rm obs}(z')}\frac{\mathrm{d}\Phi}{\mathrm{d}z'}\mathrm{d}z'\right).
\end{equation}
This deduced density fall-off can be compared with the observed fall-off $\nu_{\rm trac}^{\rm obs}(z)/\nu_{\rm trac}^{\rm obs}(z_{\rm min})$ and the goodness-of-fit between them is calculated. Although our study uses a single tracer population, the MA method does not necessarily require only a single tracer, but multiple tracers can be used. Since the MA method usually introduces a large number of parameters, an MCMC technique should be applied to explore the vast parameter space (see Sect. \ref{MCMC}). Then, as a result, probability distribution functions (PDFs) and their degrees of uncertainties are obtained for all model parameters.

\section{An assumed Galaxy model for mock observations}
\label{model}
We made no use of observational data, nor do we intend to determine the LDMD in the real Galaxy. Instead, our study performs mock observations of a tracer sample using an analytical model. To this end, we prepared an \textit{assumed} galaxy model for the purpose of creating the mock tracer data.

First, we assume that density profiles of all visible components are represented as $\rho_i^{\rm ass}\propto{\rm sech}^2\left(z/(2h_i\right))$ although this may not be suitable for gas or stellar halo components. Then, the total density profile perpendicular to the disk plane is
\begin{equation}
\label{model_density}
\rho_{\rm tot}^{\rm ass}(z)=\sum_i\rho_i^{\rm ass}(0){\rm sech}^2\left(\frac{z}{2h_i}\right) + \rho_{\rm dm},
\end{equation}
where $\rho_i^{\rm ass}(0)$, $h_i$ and $\rho_{\rm dm} $ are set arbitrarily. By integrating the Poisson equation with the boundary condition of $\frac{\partial \Phi^{\rm ass}}{\partial z}=0$ at $z=0$, the differential of this galactic potential is calculated analytically:
\begin{equation}
\label{model_potential}
\frac{\mathrm{d}\Phi^{\rm ass}}{\mathrm{d}z} = 4\pi G\left[2\sum_ih_i\rho_i^{\rm ass}(0)\tanh\left(\frac{z}{2h_i}\right) + \rho_{\rm dm}z\right].
\end{equation}
This is related to the total surface density as $\frac{\mathrm{d}\Phi^{\rm ass}}{\mathrm{d}z}= 4\pi G\Sigma^{\rm ass}(z)$, and the surface density of all observable matters is $\Sigma_{\rm s}^{\rm ass}(z)=2\sum_ih_i\rho_i^{\rm ass}(0)\tanh\left(z/(2h_i)\right)$. Next, we may integrate Eq. (\ref{notilt_Jeans}) from arbitrary $z$ to $z=+\infty$. Since $\rho_i^{\rm ass}(z)\rightarrow0$ as $z\rightarrow+\infty$,
\begin{equation}
\label{sigma_z}
\sigma_{z,i}^{2}(z) = \frac{1}{\rho_i^{\rm ass}}\int^\infty_z\rho_i^{\rm ass}\frac{\mathrm{d}\Phi^{\rm ass}}{\mathrm{d}z'}\mathrm{d}z'.
\end{equation}
Substituting Eq. (\ref{model_potential}) and $z=0$ in this equation, we can compute $\sigma_{z,i}(0)$ when a set of $\rho_i^{\rm ass}(0)$, $h_i$ and $\rho_{\rm dm} $ are given. We refer to this galaxy model as an ``assumed model".

Following \citeauthor{grl:11}, we assume that the model consists of fifteen visible components with $\rho_i^{\rm ass}(0)$ listed in Table \ref{params} that are taken from observations of \citet{fhp:06}. The LDMD is set to $\rho_{\rm dm}=0.01~{\rm M_\odot~pc^{-3}}$. In addition, we determine $h_i$ as follows: in the calculation above, $\sigma_{z,i}(0)$ are the output, and $h_i$ are the input parameters; however, $\sigma_{z,i}(0)$ can be determined by observations of nearby stars and gas, whereas $h_i$ are poorly known. Therefore, it is preferable to search for a set of $h_i$ that matches $\sigma_{z,i}(0)$ computed by Eq. (\ref{sigma_z}) to the observed values. Accordingly, we employ an MCMC method to find such a set of $h_i$. The result of the set of $h_i$ is shown in the last column of Table \ref{params}.\footnote{This MCMC calculation can also take $\rho_{\rm dm}$ as another parameter. In this case, the MCMC can estimate $\rho_{\rm dm}$ directly. However, the result prefers an extremely low LDMD $\rho_{\rm dm}\simeq0$ in this case. This may imply that the density profiles of ${\rm sech}^2\left(z/(2h_i)\right)$ cannot be applied to some components in the Galaxy.} After $h_i$ are obtained, we recompute $\sigma_{z,i}(0)$ by Eq. (\ref{sigma_z}) with the set of $h_i$. These are shown in the fifth column, indicating excellent agreement with the observations (the fourth column). We adopt $\rho_i^{\rm ass}(0)$ in the third column and $\sigma_{z,i}(0)$ in the fifth column for the assumed model described by Eq. (\ref{model_density}).
\begin{table*}
  \begin{center}
    \begin{minipage}{145mm}
      \caption{Parameters used in the assumed galaxy model. This model assumes the fifteen visible components shown in the second column, following \citeauthor{grl:11}, in which their disk mass model is taken from \citet[][the third and fourth columns]{fhp:06}. In this model, we adopt $\rho_{\rm dm}=0.01~{\rm M_\odot~pc^{-3}}$ and derive $h_i$ (the last column) with MCMC calculation. Using the set of $h_i$, the fifth column is recalculated by Eq. (\ref{sigma_z}). Our assumed model of Eq. (\ref{model_density}) uses the third and sixth columns. The uncertainties on $\sigma_{z,i}(0)$ are set to be the same as in \citeauthor{grl:11}}
      \label{params}
      $$ 
      \begin{tabular}{cccccc}
        \hline
        \noalign{\smallskip}
        \multirow{2}{*}{\#} & \multirow{2}{*}{Component} & $\rho_i^{\rm ass}(0)$ $~{\rm [M_\odot~pc^{-3}]}$  & $\sigma_{z,i}(0)$ $~{\rm [km~s^{-1}]}$ & $\sigma_{z,i}(0)$ $~{\rm [km~s^{-1}]}$ & $h_i$ $~{\rm [pc]}$ \\
           &            & (observation)            & (observation)      & (MCMC)             & (MCMC)  \\
        \noalign{\smallskip}
        \hline
        \noalign{\smallskip}
        1  & ${\rm H_2}$   & 0.021  & 4.0   & 3.80$\pm$1.0  & 32.4 \\
        2  & HI(1)         & 0.016  & 7.0   & 7.00$\pm$1.0  & 63.4 \\
        3  & HI(2)         & 0.012  & 9.0   & 9.02$\pm$1.0  & 84.7 \\
        4  & Warm gas      & 0.0009 & 40.0  & 40.1$\pm$2.0  &  564 \\
        5  & Giants       & 0.0006 & 20.0  & 20.0$\pm$2.0  &  224 \\
        6  & $M_v<2.5$     & 0.0031 & 7.5   & 7.56$\pm$2.0  & 69.2 \\
        7  & $2.5<M_v<3.0$ & 0.0015 & 10.5  & 10.7$\pm$2.0  &  103 \\
        8  & $3.0<M_v<4.0$ & 0.0020 & 14.0  & 14.1$\pm$2.0  &  144 \\
        9  & $4.0<M_v<5.0$ & 0.0022 & 18.0  & 18.1$\pm$2.0  &  197 \\
        10 & $5.0<M_v<8.0$ & 0.007  & 18.5  & 18.5$\pm$2.0  &  203 \\
        11 & $M_v>8.0$     & 0.0135 & 18.5  & 18.4$\pm$2.0  &  201 \\
        12 & White dwarfs  & 0.006  & 20.0  & 20.1$\pm$5.0  &  226 \\
        13 & Brown dwarfs & 0.002  & 20.0  & 20.0$\pm$5.0  &  224 \\
        14 & Thick disk    & 0.0035 & 37.0  & 37.7$\pm$5.0  &  519 \\
        15 & Stellar halo  & 0.0001 & 100.0 & 99.8$\pm$10.0 & 1891 \\
        \noalign{\smallskip}
        \hline
      \end{tabular}
      $$ 
    \end{minipage}
  \end{center}
\end{table*}

The total observable surface density of the assumed model is $\Sigma^{\rm ass}_{\rm s}|_{z=\infty}= 49.6~{\rm M_\odot~pc^{-2}}$, this agrees excellently with the observed value of $\Sigma_{\rm s}^{\rm obs}=49.4\pm4.6~{\rm M_\odot~pc^{-2}}$ by \citet{fhp:06}; however, it should be noted that $h_i$ seem to be somewhat shorter than observed scale heights of $\simeq300~{\rm pc}$ for Galactic thin-disk components \citep[e.g.,][]{gr:83,bm:98,bab:12I,bab:12II}.

\section{The mock data}
\label{data}
We created tracer stars dynamically consistent with the assumed galaxy model in Sect. \ref{model}. We assigned three-dimensional positions and velocities, $(x,y,z,v_x,v_y,v_z)$ to the tracer stars. In this model, the Sun is presumed to be located at $(x,y,z)=(0,0,0)$ and at rest.

As well as the model components, we assume the tracer density profile to be 
\begin{equation}
\label{tracer_density}
\nu_{\rm trac}(z) = \nu_{\rm trac}(0){\rm sech}^2\left(\frac{z}{2h_{\rm trac}}\right),
\end{equation}
where we set $h_{\rm trac}=200~{\rm pc}$. We assume that the tracer stars are uniformly distributed on the $xy$-plane at a height above the plane. Thus, we can place tracer stars at random in spatial coordinates according to Eq. (\ref{tracer_density}).

The vertical velocity dispersion profile can be calculated by Eq. (\ref{sigma_z}) using the assumed model:
\begin{equation}
\label{sigma_z_tracer}
\sigma_{z,\rm trac}^{2}(z) = \frac{1}{\nu_{\rm trac}}\int^\infty_z\nu_{\rm trac}\frac{\mathrm{d}\Phi^{\rm ass}}{\mathrm{d}z'}\mathrm{d}z'.
\end{equation}
Motions of the tracers in the Galactic radial ($x$-) and azimuthal ($y$-) directions are taken from the observational fitting functions of \citet{bis:10}. The radial velocity dispersion is 
\begin{equation}
\label{sigma_x_tracer}
\sigma_{x,\rm trac}(z) = 40 + 5\left(\frac{z}{{\rm kpc}}\right)^{1.5} {\rm km~s^{-1}}.
\end{equation}
We assume that there is no meridional motion: $\overline{v_{x,\rm trac}}=0$. The mean azimuthal velocity with respect to the sun decreases with $z$,
\begin{equation}
\label{mean_y_tracer}
\overline{v_{y,\rm trac}}(z) = -19.2\left(\frac{z}{{\rm kpc}}\right)^{1.25} {\rm km~s^{-1}}.
\end{equation}
Also, the azimuthal velocity dispersion profile is given as
\begin{equation}
\label{sigma_y_tracer}
\sigma_{y,\rm trac}(z) = 30 + 3.0\left(\frac{z}{{\rm kpc}}\right)^{2.0} {\rm km~s^{-1}}.
\end{equation}
Assuming Gaussian velocity distributions with the dispersions above, we can assign a velocity vector to each tracer star. However, it should be noted that velocity distributions may not be Gaussian in the real Galaxy. In addition, the observations of \citet{bis:10} are for blue stars, whereas the tracer population should be old (red) stars.

Now that we can assign $(x,y,z,v_x,v_y,v_z)$ to the tracer stars according to the equations and assumptions above, the positions and the velocities are converted to the spherical coordinates of $(d,\theta,\phi,v_{\rm los},v_\theta,v_\phi)$, which are centered at the Sun, where $d$ is distance from the Sun, and $v_{\rm los}$ is LOSV. We applied mock observational errors to the stellar positions and velocities (see Sect. \ref{results}), then we turned the \textit{errored} positions and velocities back to the Cartesian coordinates $(x',y',z',v_x',v_y',v_z')$. Next, we picked out stars contained in a certain sampling region, then we divided the sample stars into $z$-bins; the number of the bins was set to $n_{\rm bin}=10$ in this study. The $z$-bins have the equal widths of $(z_{\rm max}-z_{\rm min})/n_{\rm bin}$, which are centered at $z_j$. Finally, we obtained profiles of $\nu_{\rm trac}^{\rm obs}(z_j)$ and $\sigma_{z,\rm trac}^{\rm obs}(z_j)$ affected by the mock errors. We hereafter set $z_{\rm min}=0.2~{\rm kpc}$ and $z_{\rm max}=1.2~{\rm kpc}$ unless otherwise stated.

\section{The MCMC method}
\label{MCMC}
The MA method is combined with the MCMC technique to explore the vast parameter space. We introduce fitting errors of the tracer density bins as
\begin{equation}
\label{chi_bin}
\chi_\nu^2 = \frac{1}{n_{\rm bin}}\sum^{n_{\rm bin}}_{j=1}\left[\frac{\nu^{\rm ded}_{\rm trac}(z_j)/\nu^{\rm ded}_{\rm trac}(z_{\rm min})-\nu_{\rm trac}^{\rm obs}(z_j)/\nu^{\rm obs}_{\rm trac}(z_{\rm min})}{\epsilon\nu_{\rm trac}^{\rm obs}(z_j)/\nu_{\rm trac}^{\rm obs}(z_{\rm min})}\right]^2,
\end{equation}
and an error of the surface density of all observable matters as
\begin{equation}
\label{chi_surf}
\chi_\Sigma^2 = \left[\frac{\Sigma^{\rm ass}_s(z_{\rm max})-\Sigma_s(z_{\rm max})}{\epsilon\Sigma_{s}(z_{\rm max})}\right]^2,
\end{equation}
where $\epsilon$ is a constant that mimics uncertainty in actual observations (e.g., stellar mass estimation error). Throughout this study, we set $\epsilon=0.1$. Generally, the result does not depend too much on $\epsilon$. $\Sigma^{\rm ass}_s(z_{\rm max})$ is $47.1~{\rm M_\odot~pc^{-2}}$ in the assumed model. We define the total error of the fittings to be
\begin{equation}
\label{chi_tot}
\chi_{\rm tot}^2 = \left(1-w\right)\chi_\nu^2 + w\chi_\Sigma^2,
\end{equation}
where $w$ is a weight of the surface density error. Following G12, we set $w=0.1$.\footnote{In G12, $w=0.1$ was adopted although they did not introduce the parameter explicitly: they used nine bins of a tracer density profile and a surface density.} We confirmed that our results do not depend on the value of $w$ in the range between $0$ and $0.5$.

We used a Metropolis-Hasting algorithm for our sampling scheme and ran the MCMC chains of 50,000 steps: however, the first 5,000 steps were excluded as a burn-in period. Parameter ranges surveyed in our study are from $0$ to $0.2~{\rm M_\odot~pc^{-3}}$ for $\rho_{\rm dm}$, $\pm10~{\rm \%}$ and $\pm25~{\rm \%}$ for $\rho_i^{\rm ass}(0)$ of stellar and gas components, respectively. The ranges for $\sigma_{z,i}(0)$ are listed in Table \ref{params} and are much wider than actual observational uncertainties \citep{hf:00}. The medians of resulting PDFs were used as the best-fit values. The 90 \% confidence levels (intervals between the 5th and 95th percentiles) were used as ranges of uncertainties of the calculation.

\section{Results}
\label{results}
\subsection{Application to the analytical solutions}
\label{analytical_solution}
Before adapting the MA method to the mock observation data, we directly substituted the analytical velocity dispersion profile of Eq. (\ref{sigma_z_tracer}) into Eq. (\ref{ded_tracer}) and ran the MCMC by fitting with the analytical density of Eq. (\ref{tracer_density}). In this case, we can investigate intrinsic systematic biases of the MA method. 

\begin{figure}
  \includegraphics[width=\hsize]{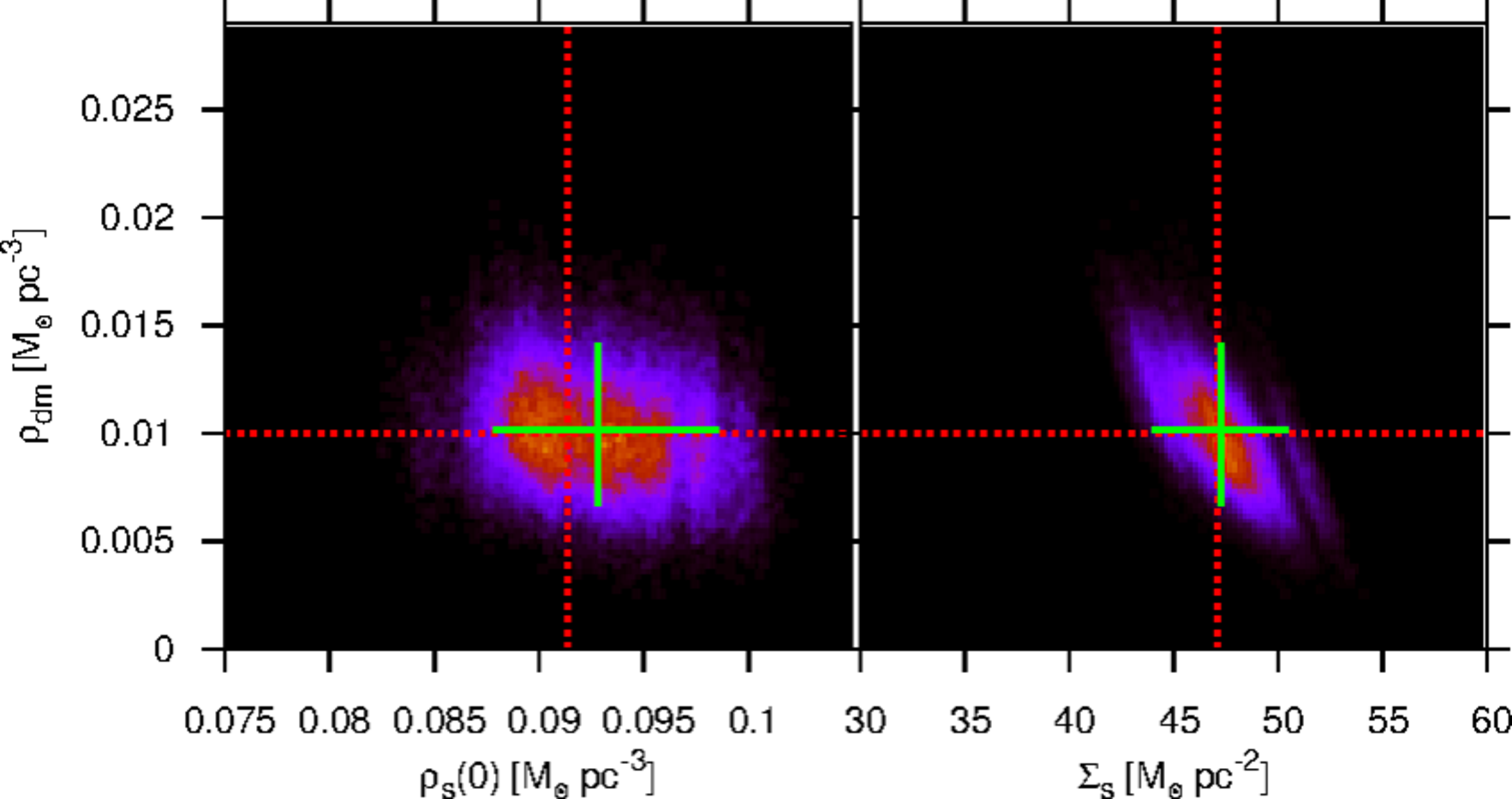}
  \caption{PDFs of LDMD, local and surface densities of all visible matters. Here, we adopt the analytical solutions of the tracer density and velocity dispersion profiles. The left and right panels illustrate the PDFs of LDMD v.s. local and surface densities of the visible matters, respectively. The brighter regions mean higher probabilities. The green solid lines indicate the ranges of the 90 \% confidence levels, and their intersection points are the medians. The red dotted lines indicate the true values in the assumed model.}
  \label{A_NoE_NoSufCon_0_contours}
\end{figure}
The result is shown in Figure \ref{A_NoE_NoSufCon_0_contours}. We can see that the LDMD is accurately determined. From this result, we expect that the isothermality assumed in Eq. (\ref{Jeans_integ}) for the deduced model does not lead to significant systematic biases (see also Appendix \ref{appendix1}) and that the MA method would be capable of determining the LDMD accurately if there are no observational errors and a sample size is large enough. We find, however, that $\rho_i(0)$ and $\sigma_{z,i}(0)$ of each component are largely degenerate with one another and not particularly well constrained.

\subsection{Application to the mock observation data}
\label{applying_mock_data}
\begin{figure}
  \includegraphics[width=\hsize]{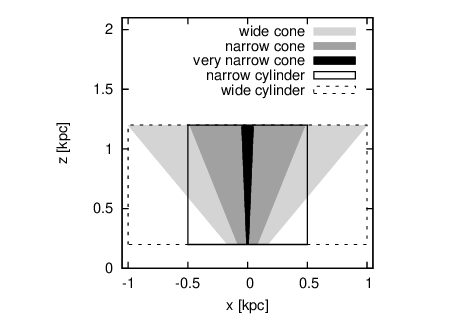}
  \caption{Sampling regions for the tracer stars. These are axisymmetric.}
  \label{regions}
\end{figure}
We generated the mock tracer population following the procedure described in Sect. \ref{data} and \textit{observed} the sample stars contained in various regions. We tested five sampling regions: ``narrow cone", ``very narrow cone", ``wide cone", ``narrow cylinder", and ``wide cylinder". These cone regions are directed to the $z$-axis ($\theta=0$) and contain stars within $\theta<22^\circ$ (narrow), $2.4^\circ$ (very narrow) and $40^\circ$ (wide). The cylinders cover the regions of $R=\sqrt{x^2+y^2}<0.5~{\rm kpc}$ (narrow) and $<1.0~{\rm kpc}$ (wide). Figure \ref{regions} delineates these sampling regions.

\subsubsection{Sample sizes}
\label{Nsample}
If tracer stars are insufficient in their sample size, statistical uncertainty may lead to an ill-determined LDMD. We investigated the sample size required to determine the LDMD precisely. For this reason, we assumed no observational errors here. In this case, since our tracer model assumes horizontally uniform distribution and kinematics, the widths of the sampling regions do not change the results. The cylinder sampling region contains more stars at lower $z$ since the stellar density decreases with $z$, whereas the cone region can cover wider areas at higher $z$. Therefore, if the sample size is fixed, samples in the cylinder and the cone can be statistically reliable in low and high-$z$ regions, respectively.

\begin{figure}
  \includegraphics[width=\hsize]{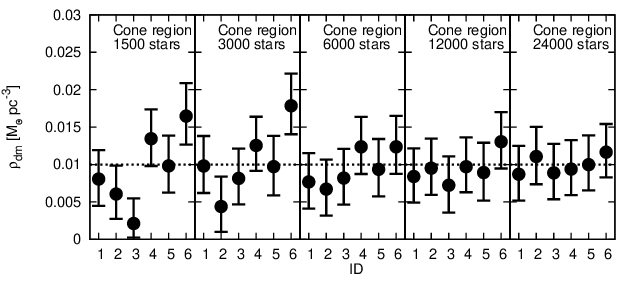}
  \includegraphics[width=\hsize]{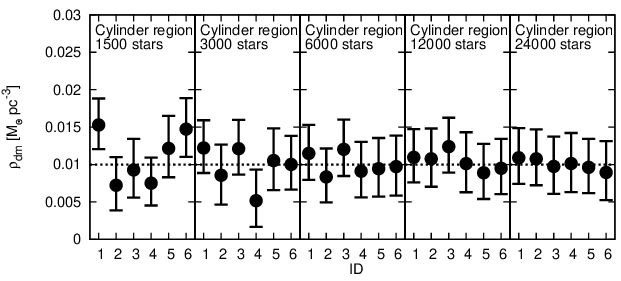}
  \caption{Results of derived LDMDs with various sample sizes. For each sample size, we performed the same computation six times with samples generated by different random seeds. The points and the error bars indicate the best-fit values and the 90 \% confidence levels, respectively. The horizontal dotted line indicates the true LDMD, $\rho_{\rm dm}=0.01~{\rm M_\odot~pc^{-3}}$.}
  \label{sampleN}
\end{figure}

We performed the calculations using the MA method with tracer samples of 1,500--24,000 stars. Our results in the cases of cone and cylinder regions are shown in Figure \ref{sampleN}. For each case, we ran the same computation six times using different random seeds in generating the samples. For the cone and cylinder sampling regions, the resulting LDMDs with $\le3,000$ stars are not precise, and the true LDMDs are often out of the error ranges. The samples of $\ge6,000$ stars seem sufficient for determining the LDMD within the errors in all runs. Therefore we consider that the minimum sample size to determine the LDMD would be approximately $\sim6,000$ stars.

\subsubsection{Distance errors}
\label{distance_error}
One of the great advantages of astrometric observations is that systematic errors are generally very small \citep[e.g.,][]{perryman:97,bpl:05,p:12}. For this reason, we examined the influence of distance-dependent random errors of parallax measurements. We assumed no systematic errors ---in other words, the mock observations are accurate but imprecise--- and no errors on stellar position measurements ($\theta$ and $\phi$); additionally, we excluded dust extinction and binary stars.

To begin with, we simply analyzed how the precisions of astrometric measurements vary with the distances of stars. Let $\varpi$ and $\mu_\theta$ be parallax and proper motion in the $\theta$-direction of a star, respectively. Except for bright sources, the uncertainty of the parallax is thought to be dominated by photon statistics to determine an image centroid \citep[e.g.,][]{bpl:05}. In this case, the parallax error $\varepsilon_{\varpi}$ varies with the observed stellar flux $f$ as $\varepsilon_{\varpi}\propto1/\sqrt{f}$. Since $f\propto d^{-2}$, $\varepsilon_{\varpi}\propto d$. If the source is bright enough to obtain a sufficient number of photons and the photons do not saturate the image, the parallax error is expected to be independent of the distance. Accordingly, we can formulate the parallax measurement error as $\varepsilon_{\varpi}\propto d^\alpha$, where $\alpha=0$ or $1$. Moreover, the parallax itself decreases as distance increases: $\varpi=1/d$. Thus, the fractional parallax error depends on the distance as $\varepsilon_{\varpi}/\varpi\propto d^{\alpha+1}$ \citep[e.g.,][]{bj:09}. Since we can equate this fractional parallax error with the fractional distance error (FDE), we formalize as 
\begin{equation}
\varepsilon_{\rm FDE}=A\left(\frac{d}{\rm kpc}\right)^{\alpha+1},
\label{FDE_eq}
\end{equation}
where $A$ is a parameter that corresponds to the precisions of parallax measurements at $d=1~{\rm kpc}$ in milliarcsecond (mas) units. We applied this FDE with Gaussian probability distributions, the dispersion of which is $\varepsilon_{\rm FDE}$ to the mock tracer stars. In addition, since astrometric measurements determine transverse velocities as $v_\theta=d\times\mu_\theta$, the velocities are also subject to the distance error as $v_\theta'=v_\theta\frac{d'}{d}$, where $v_\theta'$ and $d'$ are errored transverse velocities and distances. However, it should be noted that the formulation above may be simplistic. $\alpha$ would not be a constant but varies with $d$ in actual observations: when $\alpha=1$, the parallax uncertainties cannot be zero at $d=0$ but are limited to calibration errors and/or instrument stability at a certain distance. Additionally, for conical sampling regions with fixed angles, observations may suffer from Lutz-Kelker effect, which biases probability distribution of parallaxes toward low values by a geometrical effect and can lead to systematic errors (\citealt{lk:73,bm:98}). In this study, we ignored this effect. In this subsection, we examine the influence on determining the LDMD by the distance error alone; therefore, we ignore the errors on proper motions and LOSVs here.

\begin{figure*}
  \begin{minipage}{0.5\hsize}
    \begin{center}
      \includegraphics[width=\hsize]{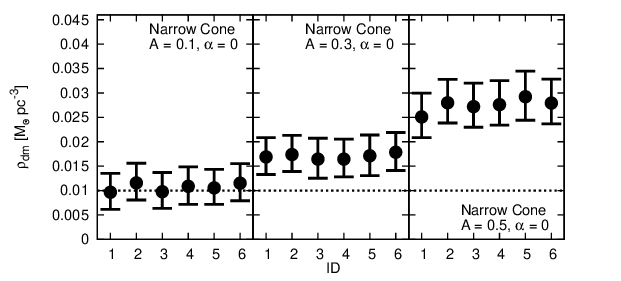}
      \includegraphics[width=\hsize]{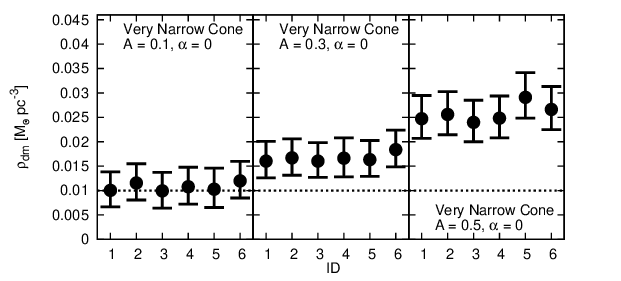}
      \includegraphics[width=\hsize]{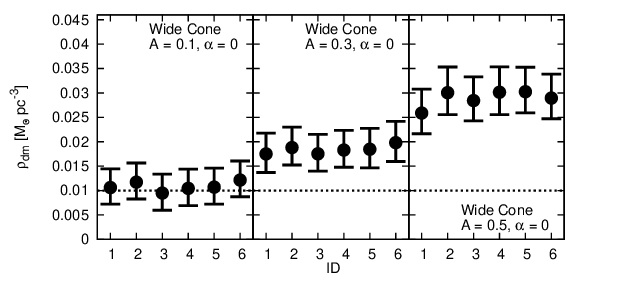}
      \includegraphics[width=\hsize]{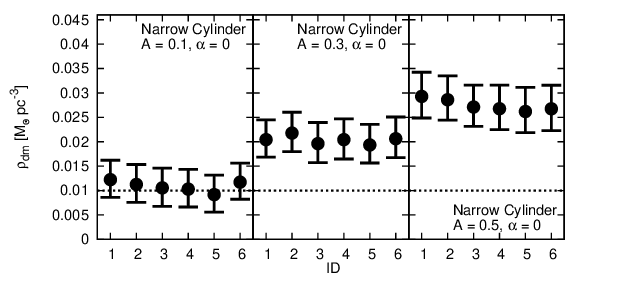}
      \includegraphics[width=\hsize]{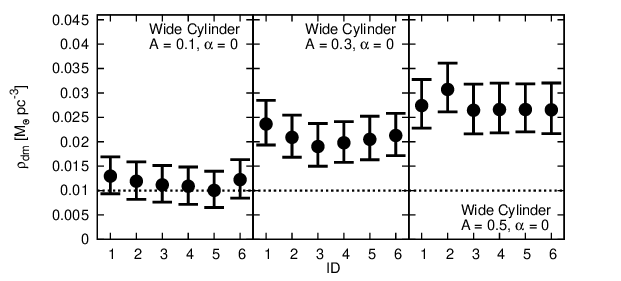}
    \end{center}
  \end{minipage}
  \begin{minipage}{0.5\hsize}
    \begin{center}
      \includegraphics[width=\hsize]{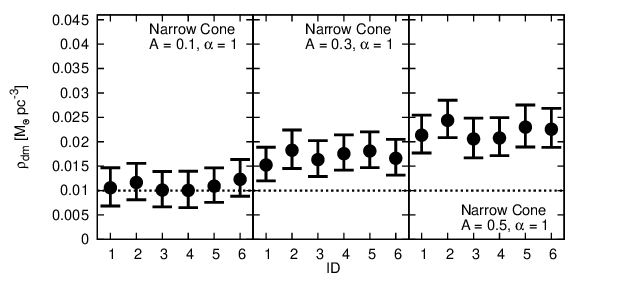}
      \includegraphics[width=\hsize]{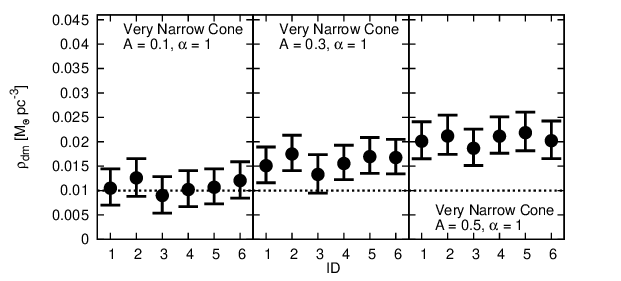}
      \includegraphics[width=\hsize]{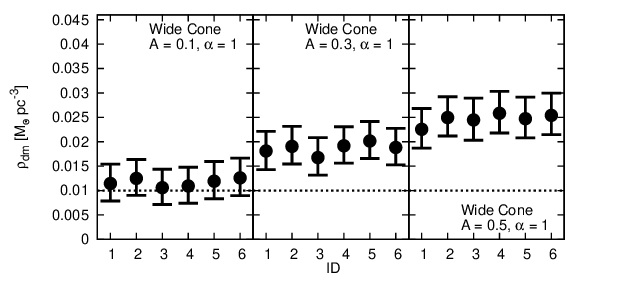}
      \includegraphics[width=\hsize]{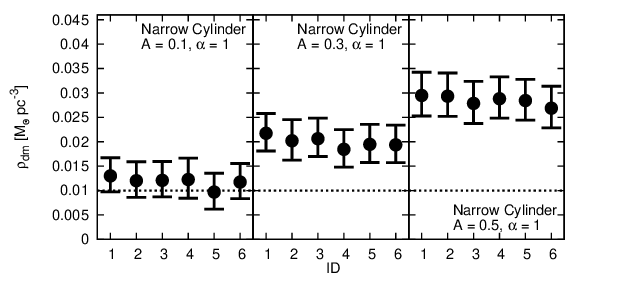}
      \includegraphics[width=\hsize]{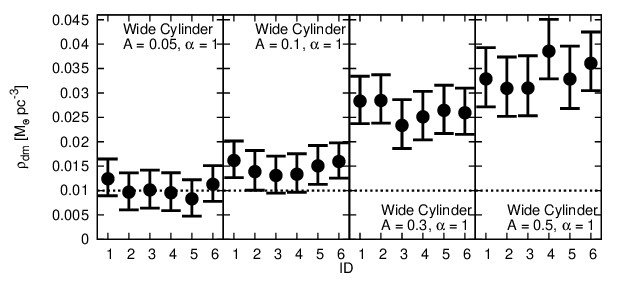}
    \end{center}
  \end{minipage}
  \caption{Influence of the astrometric distance errors on determining the LDMD. The left and right panels indicate the results of adopting $\alpha=0$ and $\alpha=1$, respectively. In each case, we conducted the same computation with different tracer samples generated in the same conditions (${\rm ID}=1$--$6$). The horizontal dotted line indicates the true LDMD. For the wide cylinder with $\alpha=1$ (the right bottom panel), we additionally show the result of adopting $A=0.05$.}
  \label{FDE}
\end{figure*}
Figure \ref{FDE} shows our results of LDMDs determined for various $A$, $\alpha$ and the sampling regions. Our results obviously demonstrate that the distance errors cause systematic overestimation of the LDMD. $\rho_{\rm dm}$ is significantly overestimated beyond the error ranges when $A\geq0.3$ in all cases except the wide cylinder with $\alpha=1$. The setting of $A=0.1$  corresponds to the parallax measurement precision of $0.1~{\rm mas}$ at $d=1~{\rm kpc}$. Only for the wide cylinder with $\alpha=1$, the required precision seems to be $0.05~{\rm mas}$.

\begin{figure}
  \includegraphics[width=\hsize]{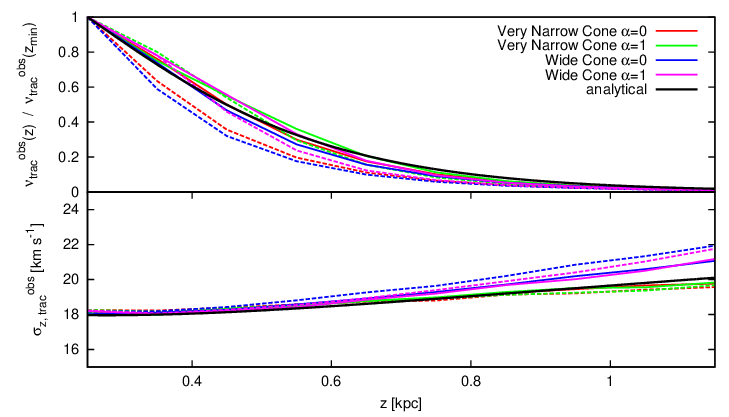}
  \includegraphics[width=\hsize]{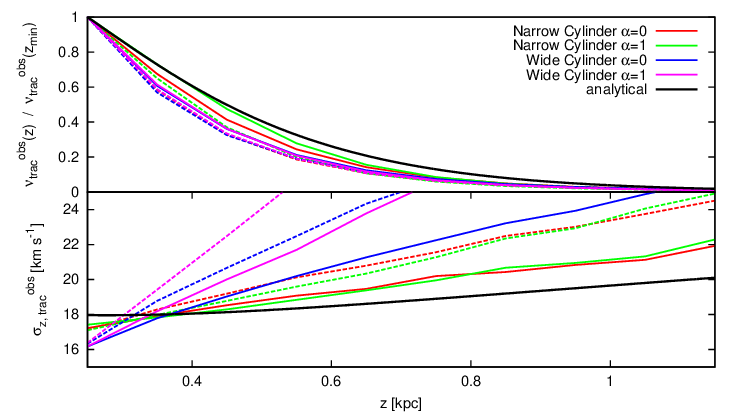}
  \caption{Changes of density fall-off and vertical velocity dispersion profiles of the tracers. Here we take into account the distance errors only. The line colors indicate results of different sampling regions and $\alpha$. The solid and dashed lines indicate results of $A=0.3$ and $0.5$, respectively. Here we used mock data of one million tracer stars in all cases.}
  \label{FDE_changes}
\end{figure}
Figure \ref{FDE_changes} shows how the distance errors deteriorate the reproducibility of these profiles. The density fall-offs are similarly affected in both the cone and the cylinder sampling regions. In the top panel, the velocity dispersion profile is hardly changed for the very narrow cone; however, the LDMD is significantly overestimated even in this case (Figure \ref{FDE}). Therefore, it can be said that the overestimation can be caused solely by the change of density fall-off; however, it can also be caused solely by the change of velocity dispersion (see Sect. \ref{LoS_error}). The changes of velocity dispersion profiles are more serious for cylinders than in the cones. This is due to inclusion among tracer stars in the cylinder regions stars whose $\theta$ are large in low-$z$ regions. The transverse velocities of these stars have more information about $v_z$ than their LOSVs.

\subsubsection{Proper motion errors}
\label{PM_error}
Proper motion uncertainties vary with distance in the same way as parallaxes in Eq. (\ref{FDE_eq}), we therefore formalize the proper motion error in the $\theta$-direction as
\begin{equation}
\varepsilon_\mu=B\left(\frac{d}{\rm kpc}\right)^\alpha {\rm mas~yr^{-1}},
\label{PM_eq}
\end{equation}
where $B$ corresponds to the precisions of proper motion measurements at $d=1~{\rm kpc}$ in the unit of ${\rm mas~yr^{-1}}$. We assumed Gaussian errors with the dispersion of $\varepsilon_\mu$ and added the errors to the true proper motions. Errored transverse velocities are $v_\theta'=d'\times\mu_\theta'$, where $\mu_\theta'$ is the errored proper motion.

\begin{figure*}
  \begin{minipage}{0.5\hsize}
    \begin{center}
      \includegraphics[width=\hsize]{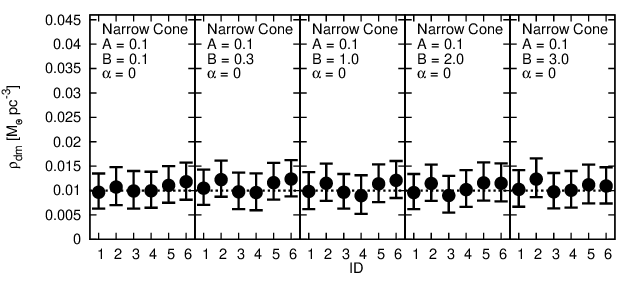}
      \includegraphics[width=\hsize]{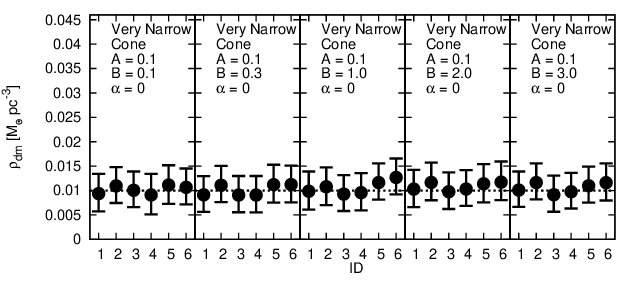}
      \includegraphics[width=\hsize]{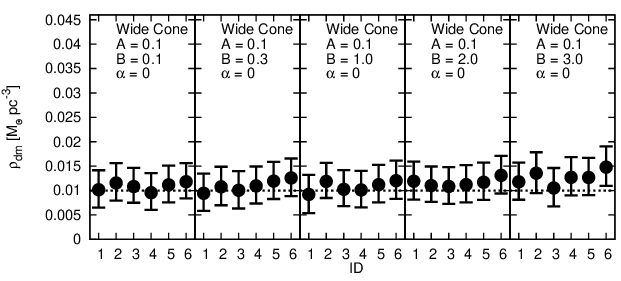}
      \includegraphics[width=\hsize]{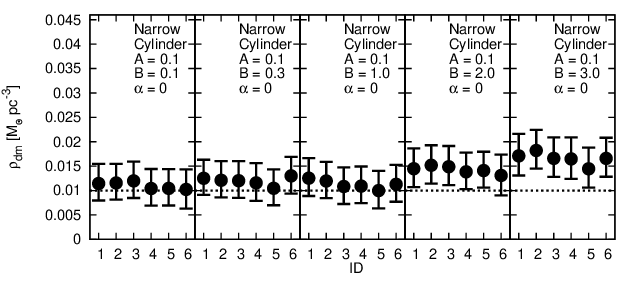}
      \includegraphics[width=\hsize]{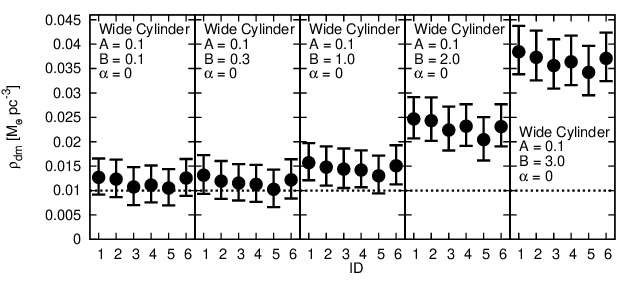}
    \end{center}
  \end{minipage}
  \begin{minipage}{0.5\hsize}
    \begin{center}
      \includegraphics[width=\hsize]{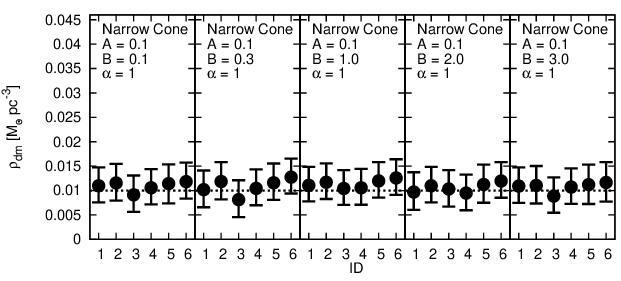}
      \includegraphics[width=\hsize]{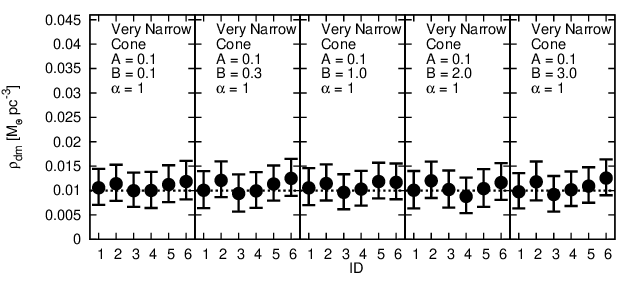}
      \includegraphics[width=\hsize]{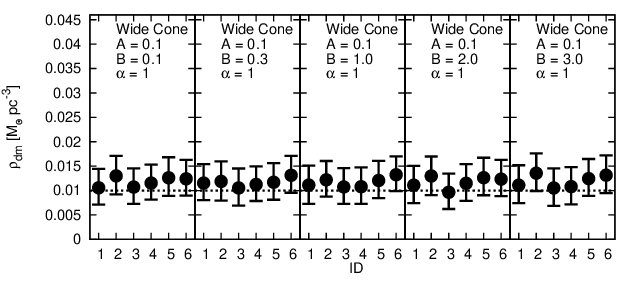}
      \includegraphics[width=\hsize]{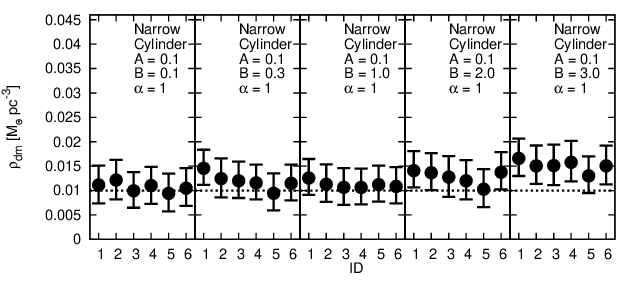}
      \includegraphics[width=\hsize]{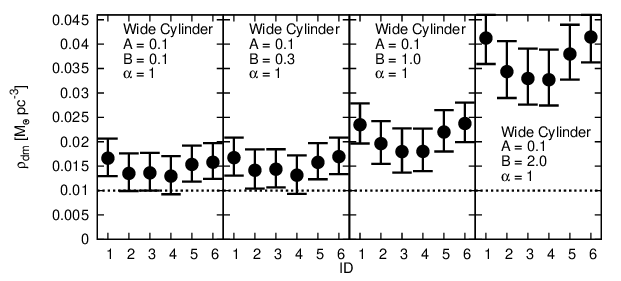}
    \end{center}
  \end{minipage}
  \caption{Influence of the proper motion errors on determining the LDMD. The left and right panels indicate the results of adopting $\alpha=0$ and $\alpha=1$, respectively. In all cases, we set the magnitude of distance error to be $A=0.1$. In each case, we conducted the same computations with different tracer samples generated from the same conditions (${\rm ID}=1$ -- $6$). The horizontal dotted line indicates the true LDMD. For the wide cylinder with $\alpha=1$ (the right bottom panel), we omitted the result of adopting $B=3.0$ since the derived LDMDs become extremely high.}
  \label{PE}
\end{figure*}
In most cases in Figure \ref{FDE}, the LDMD is overestimated by the distance errors when $A\ge0.3$; therefore, we here set $A=0.1$ in all cases. Figure \ref{PE} shows our results of adopting various $B$, $\alpha$, and sampling regions. In the figure, the overestimation is hardly visible where the cone sampling regions were used, even when $\beta=3.0$. However, a slight overestimation can be seen for the narrow cylinder when $B\ge2.0$, and significant overestimation occurs for the wide cylinder when $B\ge1.0$. Again, this is because the cylindrical regions include stars with larger $\theta$ in lower $z$ regions. Therefore, sampling in the cylindrical regions seems to include a risk of tracer kinematics becoming susceptible to the proper motion errors. Moreover, the magnitude of the overestimate seems to strongly depend on the widths of the cylinders. Accordingly, we suggest that the shapes and widths of sampling regions be chosen carefully. The wide cylinder would be useful to gather more stars in low-$z$ regions; however, if the precision of proper motion measurements is insufficient, the tracer sampling should be conducted in a conical region.

\begin{figure}
  \includegraphics[width=\hsize]{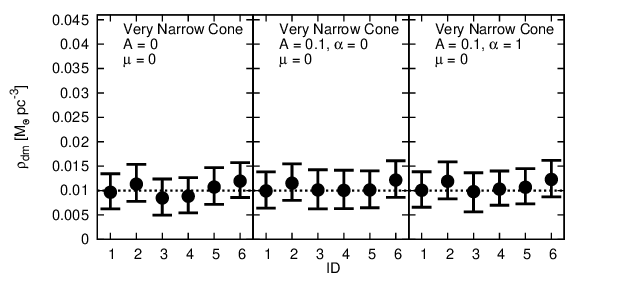}
  \includegraphics[width=\hsize]{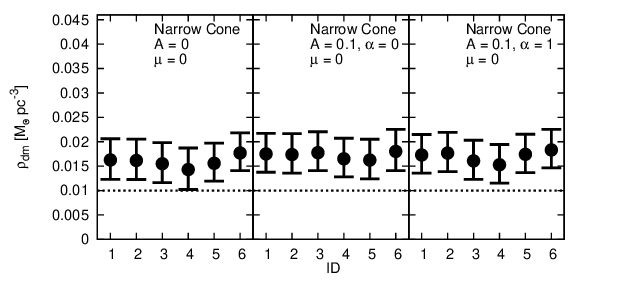}
  \caption{LDMDs determined for ignored proper motion measurement. Here we set $\mu_\theta=0$ for all tracer stars.}
  \label{DisP}
\end{figure}
From the result above, it may be expected that the proper motion measurements are not indispensable for deriving the LDMD with the MA method if the sampling region is a cone. Distances to stars can be measured not only by astrometry, but also by other observations (e.g., photometric distances). Hence, if the LDMD can be determined without proper motions, we do not necessarily need astrometry. Accordingly, we examined the cases of ignoring the proper motion, where we set $\mu_\theta=0$ for all tracers. Figure \ref{DisP} shows our results for the very narrow and narrow cone sampling regions. The LDMD can be accurately determined without proper motions if the sampling region is the very narrow cone. On the other hand, neglecting the proper motion measurements causes overestimation for the narrow cone. These results do not change even if small distance errors with $A=0.1$ are applied. Accordingly, it can be said that if tracer stars are observed in a sufficiently narrow conical region, proper motion measurements are not necessary. However, it may be difficult or impossible for such a narrow cone to contain a sufficient number of tracer stars; our result in Figure \ref{sampleN} suggests that the required sample size is approximately 6,000 stars.

\subsubsection{Line-of-sight velocity errors}
\label{LoS_error}
Since we cannot obtain LOSVs of stars from astrometry alone and need independent spectroscopic observations, observational errors on LOSVs have no relation to the distance and proper motion errors. As well as the functional forms of the astrometric errors we introduced with Eq. (\ref{FDE_eq}) and (\ref{PM_eq}), here we formalize the LOSV error as
\begin{equation}
\varepsilon_{losv}=C\left(\frac{d}{\rm kpc}\right)^\beta {\rm km~s^{-1}},
\label{LOSV_eq}
\end{equation}
where $C$ and $\beta$ are parameters. $C$ corresponds to the LOSV uncertainties at $d=1~{\rm kpc}$ in the unit of ${\rm km~s^{-1}}$. If the quality of the spectroscopic observations is good enough, the LOSV precisions would be limited to wavelength resolutions and/or calibration errors of the observations: precision floors. In this case, the LOSV precisions can be assumed to be independent of distances to stars, which correspond to $\beta=0$. Otherwise, we assume that the LOSV errors depend on the signal-to-noise ratios of the stars. If these are dominated by photon statistics, the LOSV errors can be ideally assumed to vary with stellar flux as $\varepsilon_{losv}\propto1/\sqrt{f}\propto d$. This case corresponds to $\beta=1$. However, the LOSV errors in actual observations are highly complicated \citep[e.g.,][]{k:04,bpl:05}. We therefore examined cases of $\beta=0$--$3$; in actual LOSV measurements, $\beta$ seems to be in the range of $2$--$3$ \citep[e.g.,][]{prusti:12}. However, again it should be noted that this formalization is simplistic. Even if it is the case of $\beta\ge1$, the uncertainties are limited to the precision floors at $d\simeq0$. Besides, the uncertainties depend on stellar populations. We assumed no astrometric errors here: $A=B=0$. In this case, the tracer density is not affected by the observational errors.

\begin{figure*}
  \begin{minipage}{0.5\hsize}
    \begin{center}
      \includegraphics[width=\hsize]{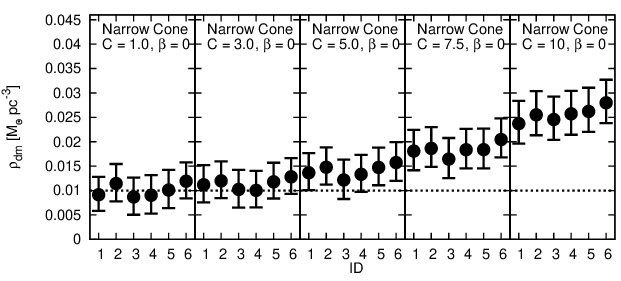}
      \includegraphics[width=\hsize]{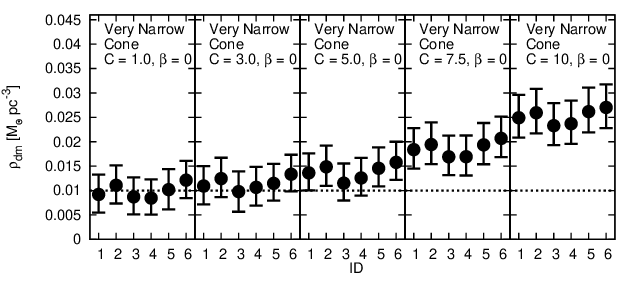}
      \includegraphics[width=\hsize]{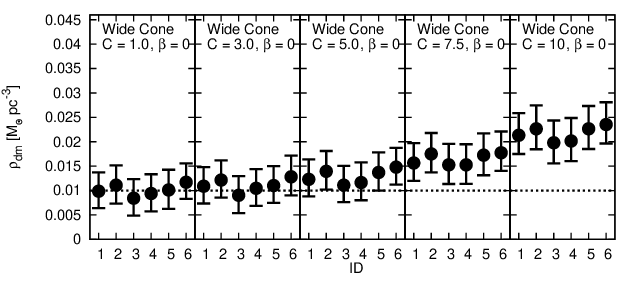}
      \includegraphics[width=\hsize]{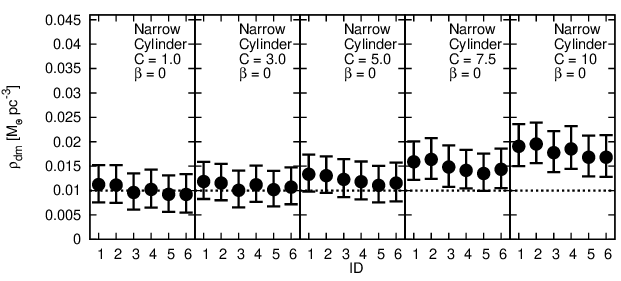}
      \includegraphics[width=\hsize]{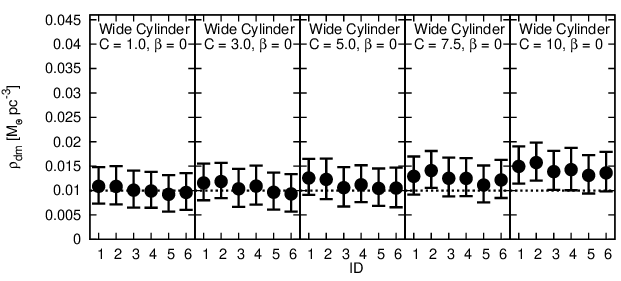}
    \end{center}
  \end{minipage}
  \begin{minipage}{0.5\hsize}
    \begin{center}
      \includegraphics[width=\hsize]{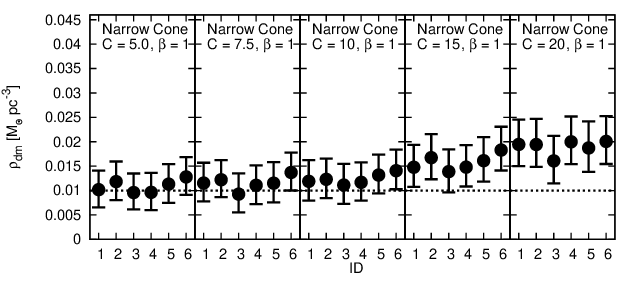}
      \includegraphics[width=\hsize]{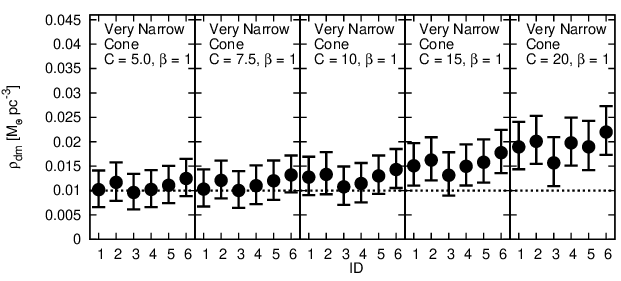}
      \includegraphics[width=\hsize]{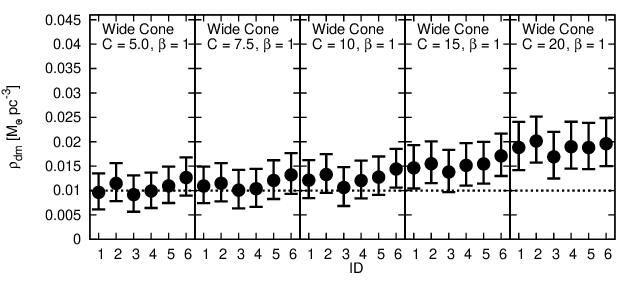}
      \includegraphics[width=\hsize]{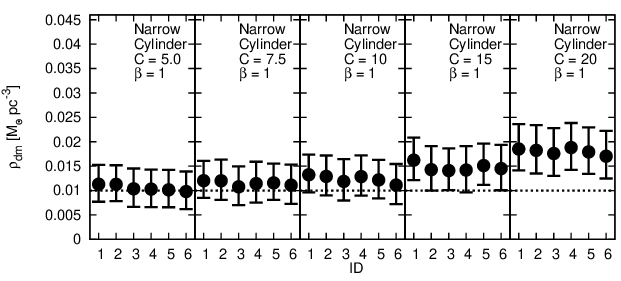}
      \includegraphics[width=\hsize]{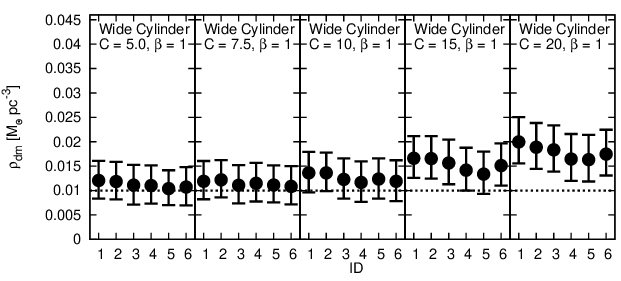}
    \end{center}
  \end{minipage}
  \caption{Influence of the LOSV errors on determining LDMD. The left and right panels indicate the results of adopting $\beta=0$ and $\beta=1$, respectively. In all cases, we assumed that there are no distance and proper motion errors: $A=B=0$. In each case, we conducted the same computations with different tracer samples generated from the same conditions (${\rm ID}=1$ -- $6$). The horizontal dotted line indicates the true LDMD.}
  \label{LOSV1}
\end{figure*}
\begin{figure*}
  \begin{minipage}{0.5\hsize}
    \begin{center}
      \includegraphics[width=\hsize]{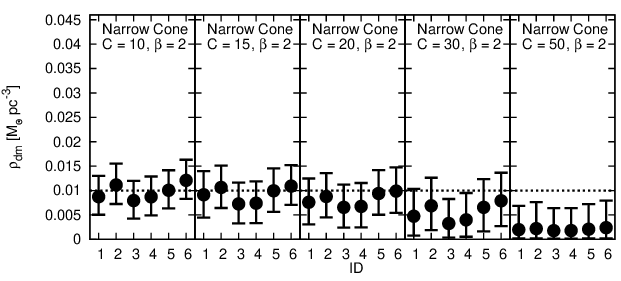}
      \includegraphics[width=\hsize]{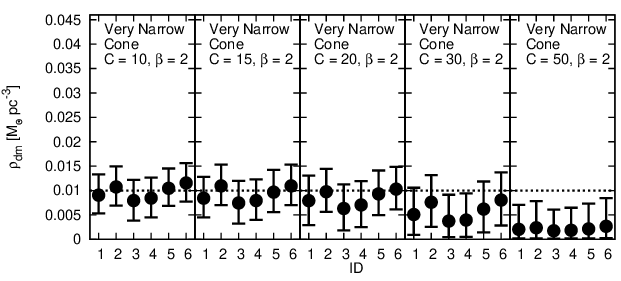}
      \includegraphics[width=\hsize]{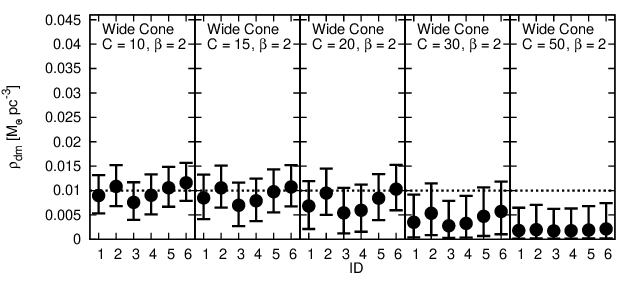}
      \includegraphics[width=\hsize]{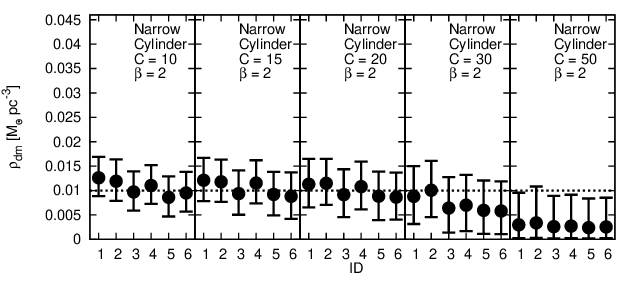}
      \includegraphics[width=\hsize]{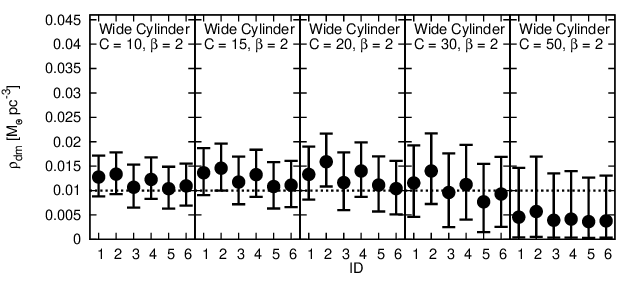}
    \end{center}
  \end{minipage}
  \begin{minipage}{0.5\hsize}
    \begin{center}
      \includegraphics[width=\hsize]{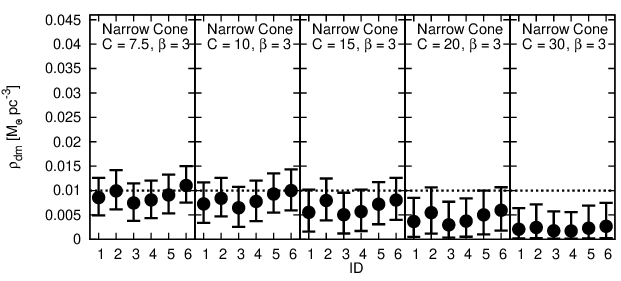}
      \includegraphics[width=\hsize]{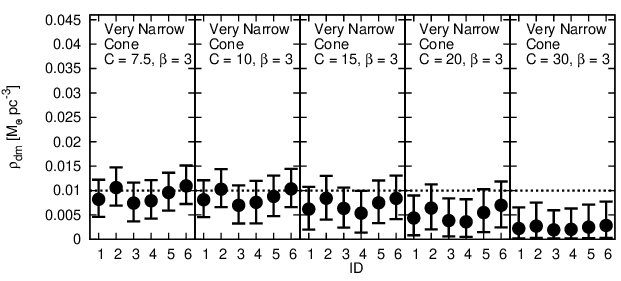}
      \includegraphics[width=\hsize]{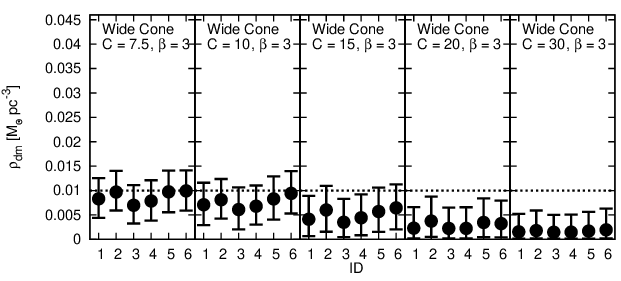}
      \includegraphics[width=\hsize]{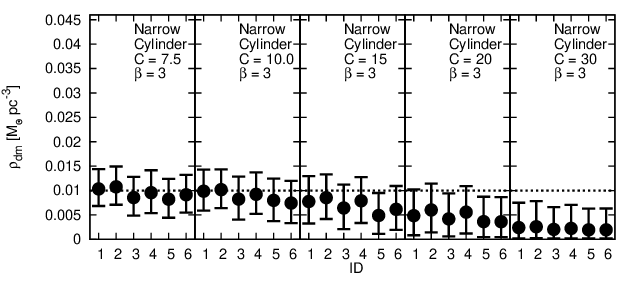}
      \includegraphics[width=\hsize]{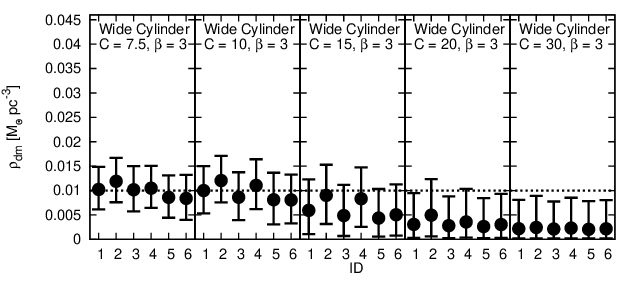}
    \end{center}
  \end{minipage}
  \caption{Same as Figure \ref{LOSV1}, but adopting $\beta=2$ (left panels) and $\beta=3$ (right panels).}
  \label{LOSV2}
\end{figure*}
Figures \ref{LOSV1} and \ref{LOSV2} show our results of adopting the various $C$, $\beta$, and sampling regions. For $\beta=0$, the conical sampling regions overestimate the LDMD when $C>3.0~{\rm km~s^{-1}}$, and the cylinders overestimate it when $C>5.0~{\rm km~s^{-1}}$ in some runs. These values of $C$ can be considered as the required precision floors of the LOSV measurements. Where $\beta=1$, all sampling regions overestimate the LDMD when the errors are $>7.5~{\rm km~s^{-1}}$ at $d=1~{\rm kpc}$ in some runs. Where $\beta=2$ and $3$, the behavior of the derived LDMDs is interesting; the LDMDs are underestimated. Where $\beta=2$, the LOSV precision required to prevent significant underestimation is approximately $20~{\rm km~s^{-1}}$ at $d=1~{\rm kpc}$ for the conical sampling regions and $30~{\rm km~s^{-1}}$ for the cylinders. Where $\beta=3$, the precision required is approximately $10~{\rm km~s^{-1}}$ at $d=1~{\rm kpc}$ for the conical sampling regions and $15~{\rm km~s^{-1}}$ for the cylinders. 

\begin{figure}
  \includegraphics[width=\hsize]{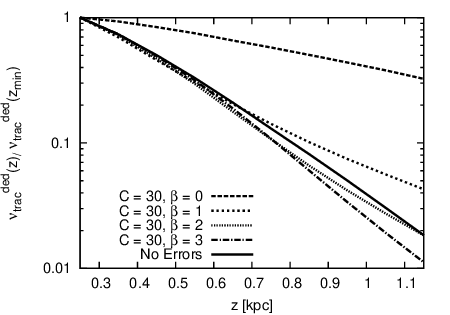}
  \caption{Tracer density fall-offs deduced by the MA method. Here we set all parameters to the true values and applied the LOSV errors only. Because distance errors are not taken into account, the deduced profile for no errors corresponds to the observed density fall-off. The mock data of one million stars were used.}
  \label{LOSV_tracer_changes}
\end{figure}
Figure \ref{LOSV_tracer_changes} shows tracer density fall-offs \textit{deduced} from Eq. (\ref{ded_tracer}) under the LOSV errors. All parameters of $\rho_i(0)$, $\sigma_{z,i}(0)$ and $\rho_{\rm dm}$ are set to the true values. This figure explains why the LOSV errors underestimate or overestimate the LDMD depending on $\beta$. Where $\beta=0$ and $1$, the MA method tends to predict thicker distributions of the tracer than when the LOSV errors are excluded; on the other hand, for $\beta=2$ and $3$, the MA method predicts thinner distributions. Generally, higher LDMDs make the tracer distribution thinner, while lower LDMDs yield thicker tracer distribution. Therefore, in the cases of $\beta=0$ and $1$, high LDMDs are required to match with the observed density fall-offs; in the cases of $\beta=2$ and $3$, low LDMDs are preferred.

\section{Discussion}
\label{discussion}
\subsection{Required observational precision}
\label{comparison}

In Sect. \ref{Nsample}, we found that the sample size required to determine the LDMD with accuracy is approximately 6,000 stars. G11 have used three tracer populations in their study, and their sample sizes are 139 K giants from $z=0.2~{\rm kpc}$ to $0.7~{\rm kpc}$, 2026 A stars, and 3080 F stars in $z\le0.2~{\rm kpc}$. Although the total number of their tracer stars is similar to the required sample size we estimated, their sample of the K stars in the high-$z$ region seems to be insufficient in number. G12 used a single tracer population that consists of 2016 K stars used for the density fall-off and 580 K stars used for the velocity dispersion profile. The sample size of G12 is still smaller than the required sample size we estimated.

In Sect. \ref{distance_error}, we estimated that the required precision of the parallax measurements is approximately $\sigma_\varpi\hspace{0.3em}\raisebox{0.4ex}{$<$}\hspace{-0.75em}\raisebox{-.7ex}{$\sim$}\hspace{0.3em}0.1~{\rm mas}$ at $d=1~{\rm kpc}$ in almost all cases. Standard errors of $Hipparcos$ observations are in the range of $\sigma_\varpi=0.7$--$0.9~{\rm mas}$ even for stars brighter than ninth magnitude (mag) in a catalog published in 1997 \citep{perryman:97,p:12}.

\begin{figure}
  \includegraphics[width=\hsize]{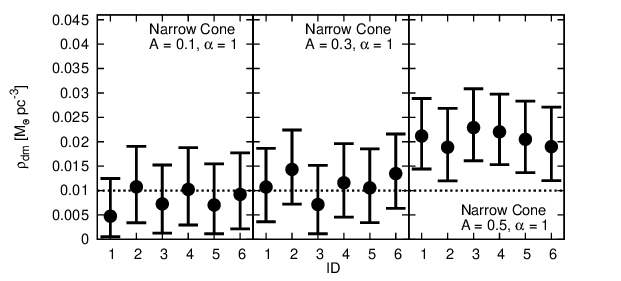}
  \includegraphics[width=\hsize]{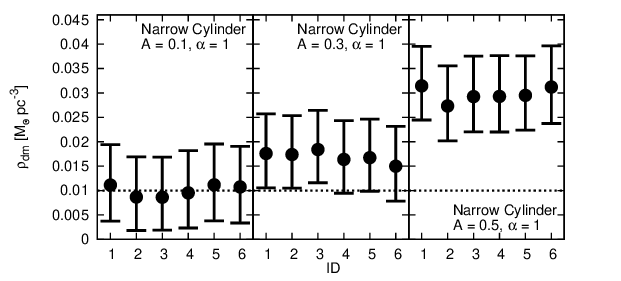}
  \caption{Influence of the distance errors on determining the LDMD for $z_{\rm max}=0.75~{\rm kpc}$ and $z_{\rm min}=0.2~{\rm kpc}$. The other settings are the same.}
  \label{shortZmax}
\end{figure}
It should be noted that our results may depend on details of our settings. For example, we arbitrarily set the scale height of our tracer to $h_{\rm trac}=200~{\rm pc}$. $z_{\rm min}=0.2~{\rm kpc}$ and $z_{\rm max}=1.2~{\rm kpc}$ are also arbitrary. It is expected that such a small sample can avoid high-$z$ regions where the observational errors become large. Figure \ref{shortZmax} shows determined LDMDs for $z_{\rm max}=0.75~{\rm kpc}$. Here, the parallax measurement errors are taken into account. Although the overestimation of the LDMD is slightly mitigated for the narrow cone sampling region, significantly high LDMDs are predicted when $A\ge0.5$. Therefore, we expect that the required parallax precisions are $\sigma_\varpi=0.1$--$0.3~{\rm mas}$.

\begin{figure}
  \includegraphics[width=\hsize]{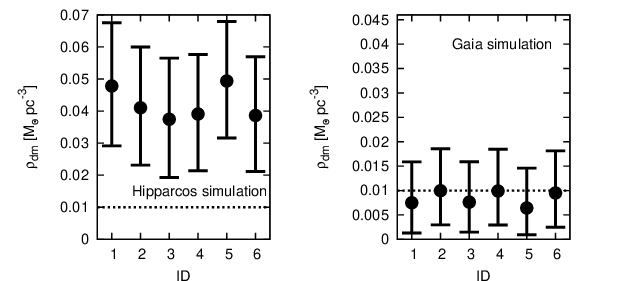}
  \caption{LDMD determinations by our simulations of \textit{Hipparcos} (left) and \textit{Gaia} (right) observations adopting Eq. (\ref{Hipprcos_FDE}) and (\ref{Gaia_FDE}), respectively. Here we ignored the proper motion and the LOSV errors and set $z_{\rm min}=0$ and $z_{\rm max}=0.75~{\rm kpc}$. The sampling regions are the narrow cylinder in both cases.}
  \label{Hipparcos_Gaia}
\end{figure}
We can readily simulate the \textit{Hipparcos} observations by modifying our FDE model of Eq. (\ref{FDE_eq}). The floor precision of parallax measurements by \textit{Hipparcos} is approximately $0.7~{\rm mas}$ \citep{perryman:97,p:12}, and the FDEs increase linearly within $d\hspace{0.3em}\raisebox{0.4ex}{$<$}\hspace{-0.75em}\raisebox{-.7ex}{$\sim$}\hspace{0.3em}100~{\rm pc}$ for A and F stars \citep{hf:00}; this floor precision corresponds to $\varepsilon_{\rm FDE}=0.07$ at $d=100~{\rm pc}$. Therefore, here we used the following FDE model instead of Eq. (\ref{FDE_eq}):
\begin{equation}
\varepsilon_{\rm FDE}=0.07\left(\frac{d}{100~{\rm pc}}\right)^{\alpha+1},
\label{Hipprcos_FDE}
\end{equation}
where $\alpha=0$ when $d<100~{\rm pc}$, and $\alpha=1$ when $d>100~{\rm pc}$. Our result for the narrow cylinder region is shown in the left panel of Figure \ref{Hipparcos_Gaia}. In this simulation, we set $z_{\rm min}=0$ and $z_{\rm max}=0.75~{\rm kpc}$. The \textit{Hipparcos} simulation clearly overestimates the LDMD. This result indicates that the \textit{Hipparcos} catalog is too imprecise to determine the LDMD with accuracy. The error ranges in these simulations are much larger than in our previous calculations. This is because of the large parallax uncertainties of \textit{Hipparcos} and the settings of $z_{\rm min}=0$ and $z_{\rm max}=0.75~{\rm kpc}$. The region near to the disk plane has little information about the LDMD, where observable densities are nearly flat and degenerate with the background DM. Moreover, because the sample size is now fixed, the inclusion of the low-$z$ region leads to poor statistics in high-$z$ region. \footnote{We included the low-z region to discuss how the transition of $\alpha$ affects the LDMD determination.}

In Sect. \ref{PM_error}, we estimated the required precision of proper motion measurements, which seems higher than $3~{\rm mas~yr^{-1}}$ at $d=1~{\rm kpc}$. For the cylindrical sampling regions, however, the required precision becomes severe: approximately $\sigma_\mu\le1~{\rm mas~yr^{-1}}$  at $d=1~{\rm kpc}$ for the narrow cylinder and $\sigma_\mu\le0.3~{\rm mas~yr^{-1}}$ for the wide cylinder.

In Sect. \ref{LoS_error}, we estimated the required precisions of LOSV measurements and found that the LOSV errors lead to either overestimation or underestimation of the LDMD, which is contingent on the distance dependence of the errors. When we applied uniform errors $\beta=0$, the required precision was found to be $\le3.0~{\rm km~s^{-1}}$ for the cylindrical sampling regions and $5.0$--$7.5~{\rm km~s^{-1}}$ for the cylinder sampling regions. The required precision in this case corresponds to required precision floors. Where the errors only weakly depend on distances, the LOSV errors can cause an overestimation of the LDMD, whereas for a strong distance dependence of the errors, the LOSV errors can cause underestimation.

In Sect. \ref{analytical_solution}, our results demonstrated that the MA method does not indicate any systematic biases and is capable of determining the LDMD with accuracy if sample size and observational precisions are sufficient. Our study, however, does not discuss the LDMDs determined by methods other than the MA method. Although G11 and G12 demonstrated that the methods of \cite{hf:00} and \citet{kg:89II,kg:89III,kg:89I} are systematically biased to underestimate the LDMD, the other methods \citep[e.g.,][]{zrv:12} should also be assessed carefully. Some studies have used data of photometric distances \citep[e.g., G12;][]{zrv:12}. However, our study cannot discuss the accuracy of the LDMD determined by these studies using the photometric distances. Photometric distances may have systematic errors. For example, \citet{bab:12II} have shown that stellar distances of G dwarfs determined with an isochrone of \citet{apm:09} are nearly 10 \% larger than those determined with an isochrone of \citet{isj:08} in \textit{the Sloan Digital Sky Survey} observations. Systematically large distances of tracer stars result in large scale-heights of the distributions. Such thick tracers can underestimate the LDMD if the systematic bias is significant.

\subsection{Validity of our galaxy modeling}
\label{future}
We have to mention, however, that our analytical model for generating the mock data is somewhat simplistic. For example, we assumed that the galaxy model consists of the 15 visible matter components. This modeling may not necessarily well represent the real Galaxy. \citet{bab:12I,brh:12III,bab:12II} have recently discussed that the Galactic disk is composed of a lot of continuous ``mono-abundance sub-populations" and that there are no distinct thin/thick disks. If this is the case, the stellar classification provided in \citet[][Table \ref{model} in this paper]{fhp:06} may not be suitable. In addition, as an observational difficulty, even if data are accurate and precise enough, it may be laborious to pick out kinematically homogeneous tracer stars and to verify the homogeneity. Contaminated tracer samples can lead to ill-determined LDMDs \citep[e.g.,][]{kg:89II,kg:89III}. This also needs reliable stellar population synthesis theory and spectroscopic observations.

The tracer in our study was modeled to have uniform density and velocity distributions in a $z$-plane. But, stellar density and velocity dispersions in the Galaxy vary in the radial direction with scale radii of $\sim3.0~{\rm kpc}$ \citep[e.g.,][]{lf:89,bab:12II}.\footnote{\citet{brh:12III} have observationally shown that the scale radii of vertical velocity dispersions are as long as $\sim7.1~{\rm kpc}$.} Therefore, if tracer-sampling regions are too wide, the kinematic homogeneity for the tracer can be broken in actual observations (see G11).

If our modeling can represent the real Galaxy well, our mock observations may be used to calibrate erroneously determined LDMDs. An advantageous point of astrometric observations is that observational uncertainties are already known. Therefore, even if sufficiently precise data are not available, our method can estimate magnitudes of under/overestimation of the ill-determined LDMDs by applying the known astrometric errors to the mock data.

\subsection{Toward future astrometric observations}
\label{future}
Our results have shown that the MA method requires a high degree of astrometric precision to determine the LDMD, which still cannot be achieved by present astrometric observations. The near-future astrometric satellite \textit{Gaia}, however, will significantly exceed \textit{Hipparcos}. \textit{Gaia} is expected to observe the complete sample of all stars brighter than $20~{\rm mag}$ with end-of-mission parallax precisions of $\simeq0.01~{\rm mas}$ at $V=10~{\rm mag}$,\footnote{$V$ denotes Johnson $V$ magnitude} $0.01$--$0.03~{\rm mas}$ at $V=15~{\rm mag}$ and up to $0.1$--$0.35~{\rm mas}$ at $V=20~{\rm mag}$ \citep[e.g.,][]{bpl:05,bj:09,jgc:10,prusti:12}; proper motion precisions in the unit of ${\rm mas~yr^{-1}}$ are comparable with the parallax precisions. A frequently used tracer population are K dwarfs, whose absolute magnitude is $V\sim7~{\rm mag}$ \citep{bm:98}. This means that K stars at $d=1~{\rm kpc}$ have a brightness of $V\sim17~{\rm mag}$. Therefore, these stars are expected to be observed with precisions of $<0.1~{\rm mas}$. Thus \textit{Gaia} will be able to achieve the required astrometric precision we estimated. Furthermore, \textit{Gaia} is designed to measure LOSVs of the stars simultaneously. The LOSV precisions are $10$--$17~{\rm km~s^{-1}}$ at $V=17~{\rm mag}$ with precision floors of $\simeq1~{\rm km~s^{-1}}$ \citep{bpl:05,k:04,jgc:10}, whose distance dependence apparently is $\beta=2$--$3$ \citep[from comparison with Figure 7 of ][]{prusti:12}. From comparing with Figure \ref{LOSV2}, \textit{Gaia} seems to be able to determine the LDMD within the ranges of 90 \% confidence levels although there may be a slight underestimation if $\beta=3$ and $C=15$. Accordingly, \textit{Gaia} is expected to enable determining the LDMD using the MA method although it may be better to replace the LOSV data with more precise measurements.

Here we try to simulate the $Gaia$ observations. The floor precision of the parallax measurements of $Gaia$ is $\sim0.01~{\rm mas}$, and the uncertainties increase for stars brighter than $G\sim12~{\rm mag}$\footnote{$G$ denotes magnitude in $Gaia$-band.} \citep[e.g.,][]{prusti:12}. This brightness roughly corresponds to a K-dwarf star at $d\sim100~{\rm pc}$.\footnote{$G-V\simeq-0.4$ for the K stars \citep{jgc:10}.} A parallax uncertainty of $0.01~{\rm mas}$ corresponds to $\varepsilon_{\rm FDE}=0.001$ at $d=100~{\rm pc}$. Accordingly, we assume the following FDE model for our $Gaia$ simulation:
\begin{equation}
\varepsilon_{\rm FDE}=0.001\left(\frac{d}{100~{\rm pc}}\right)^{\alpha+1},
\label{Gaia_FDE}
\end{equation}
where $\alpha=0$ when $d<100~{\rm pc}$, and $\alpha=1$ when $d>100~{\rm pc}$. The right panel of Figure. \ref{Hipparcos_Gaia} shows the determinations of the LDMD by our $Gaia$ simulation. In this simulation, we set $z_{\rm min}=0$ and $z_{\rm max}=0.75~{\rm kpc}$, and the narrow cylinder sampling region was used. This result clearly demonstrates that the MA method will be able to determine the LDMD with accuracy when the \textit{Gaia} catalog becomes available. Although the error ranges of these simulations are somewhat larger than the other calculations in this paper, this is again because of the settings of the low $z_{\rm min}$ and $z_{\rm max}$. It seems important to cover the high-$z$ region to reduce the uncertainty of the LDMD determination. However, as we showed in Sect. \ref{distance_error}, large distance errors in the high-$z$ region may cause overestimation. Moreover, the high-$z$ region may not have a sufficient number of tracer stars. Hence, we suggest that careful analyses be performed using various $z$-ranges and tracer populations even when the \textit{Gaia} catalog becomes available.

In addition to \textit{Gaia}, \textit{Nano-JASMINE} and \textit{JMAPS} are also individually planned to produce all-sky astrometric catalogs. By combining their data with the \textit{Hipparcos} and \textit{Tycho} positional data from 1991, proper motions with $0.1~{\rm mas~yr^{-1}}$ are expected to be achievable for bright stars \citep{mlh:12}. This high-precision astrometry will improve not only the LDMD determination but also our understanding of various aspects of Galactic dynamics.

\section{Conclusions}
\label{conclusions}
We scrutinized the MA method devised by G11 and G12 and carefully assessed their LDMD determinations. We created mock observational data and applied the MA method. As results, we found that the MA method is capable of determining the LDMD with accuracy if the sample size of a tracer and observational precision is sufficient. We found that the sample size must be larger than approximately 6,000 stars. Astrometric errors, however, can cause overestimation on the LDMD determination, and we estimated that the required precision is approximately $0.1$--$0.3~{\rm mas}$ for parallax measurements. Proper motion precision does not seem to be as important as the parallax precision. In addition, LOSV errors can cause either overestimation or underestimation of the LDMD determination: if the errors weakly or strongly depend on distance, the LDMD can be overestimated or underestimated, respectively.

From our results, we expect that the MA method will overestimate the LDMD when \textit{Hipparcos} data are used, because of the insufficient astrometric precision. The near-future astrometric satellite \textit{Gaia} can be expected to measure parallaxes, proper motions, and LOSVs with sufficient precisions. Our results indicate that \textit{Gaia} will enable us to determine the LDMD using the MA method.

\begin{acknowledgements}
We would like to thank the referee for a careful reading of the manuscript and useful comments. We acknowledge Taihei Yano, Shingo Kahima, Takuji Hara, Yuji Chinone, and Ken-ichi Tadaki for their helpful discussion. Numerical computations and data analyses were in part carried out on the analysis server system at the Center for Computational Astrophysics, CfCA, of National Astronomical Observatory of Japan. This research was partially supported by the Ministry of Education, Science, Sports and Culture, Grant-in-Aid for Scientific Research (A), No.23244034, 2011-2015. 
\end{acknowledgements}

\bibliographystyle{aa}
\bibliography{references}

\begin{appendix}
\section{Validity of the isothermality for the deduced model}
\label{appendix1}
The deduced galaxy models of G11 and G12 are slightly different from each other. G11 has used parametrized runs of velocity dispersions assuming a behavior similar to the observational fitting in \citet{bis:10}. G12, on the other hand, assumed that all components are isothermal. Here, we discuss the impact of these velocity dispersion profiles on the calculation of the MA method. In addition, we verify the appropriateness of the isothermal assumption used in G12 and this study.
 
The dispersion runs introduced in G11 are described by a quadratic form:
\begin{equation}
\sigma_{z,i}^2(z)=\sigma_{z,i}^2(0)\left(1+cz^2\right),
\end{equation}
where the constant of $c$ is chosen so that this function satisfies the observational fitting reported in \citet{bis:10}:
\begin{equation}
\sigma_{z,i}(z_{\rm max})=\sigma_{z,i}(0)+4\left(\frac{z_{\rm max}}{{\rm kpc}}\right)^{1.5}.
\end{equation}
\begin{figure}
  \includegraphics[width=\hsize]{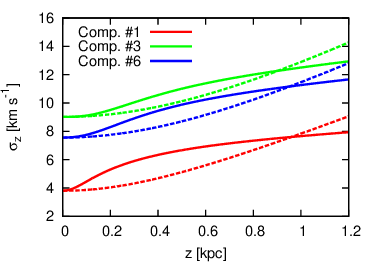}
  \includegraphics[width=\hsize]{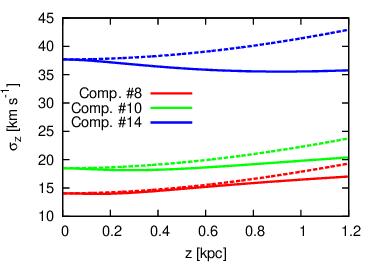}
  \caption{Velocity dispersion profiles of some model components in Table \ref{params}. The solid and dashed lines indicate analytical solutions computed with Eq. (\ref{sigma_z}) and fittings with the quadratic function, respectively.}
  \label{dispersions}
\end{figure}
Figure \ref{dispersions} shows a comparison between the quadratic functions and the analytical solutions of the Jeans equation for some model components. Although some components seem to be fitted well by the quadratic functions, some others are fitted poorly. This is because the observations of \citet{bis:10} were made for blue disk stars and/or our assumed model is artificial.

\begin{figure}
  \includegraphics[width=\hsize]{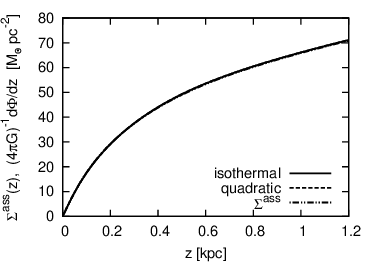}
  \includegraphics[width=\hsize]{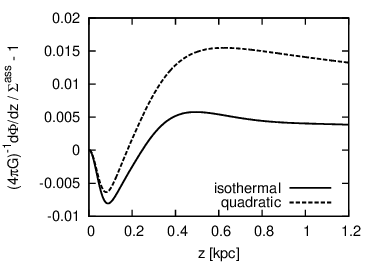}
  \caption{Comparison between surface densities deduced from the Jeans and Poisson equations $(4\pi G)^{-1}\mathrm{d}\Phi/\mathrm{d}z$ and the density profile assumed in the model, $\Sigma^{\rm ass}$. The top panel indicates the surface density profiles. Since the three lines almost overlap, they may be indistinguishable by eye. The bottom panel shows fractional differences between the deduced surface density profiles and the assumed model.}
  \label{iso-Bond}
\end{figure}
We set all parameters of $\rho_i(0)$, $\sigma_{z,i}(0)$ and $\rho_{\rm dm}$ to the true values and computed surface density profiles of the deduced models, $(4\pi G)^{-1}\mathrm{d}\Phi/\mathrm{d}z$, following the procedure of Sect. \ref{MA_model} with the quadratic functions and the isothermality. Figure \ref{iso-Bond} shows the results. In the top panel, we compare the results with the assumed density profile of Eq. (\ref{model_potential}). Since all of them are almost consistent, it can be said that both the quadratic functions and the isothermality can reproduce the assumed galactic potential well. The bottom panel shows fractional differences between the results and the assumed model. In both cases, the deduced models underestimate the surface density in $z\hspace{0.3em}\raisebox{0.4ex}{$<$}\hspace{-0.75em}\raisebox{-.7ex}{$\sim$}\hspace{0.3em}0.2~{\rm kpc}$ and overestimate in $z\hspace{0.3em}\raisebox{0.4ex}{$>$}\hspace{-0.75em}\raisebox{-.7ex}{$\sim$}\hspace{0.3em}0.2~{\rm kpc}$. However, magnitudes of the overestimation at $z_{\rm max}$ are only $\sim0.5~\%$ and $\sim1.5~\%$ for the isothermal and the quadratic dispersions, respectively. From these results, we can see that the isothermality for the deduced model hardly affects the determination of the LDMD. In addition, the runs of velocity dispersions of the model components do not seem to be important. G11 have also come to the same conclusion and mentioned the following: Although the velocity dispersion profile of the tracer directly affects the tracer density fall-off and has a significant impact on the result, uncertainties in dispersion profiles of the model components are, by contrast, marginalized out when we calculate $\rho_{\rm dm}$ and $\rho_{\rm s}(0)$, since they appear only in Eq. (\ref{Poisson3}) through Eq. (\ref{Jeans_integ}). From the above results, it is verified that assuming isothermality for the model components does not lead to an erroneous determinations of the LDMD. Although the quadratic function used in G11 is not necessarily better than the isothermal approximation, the impact of the difference between them seems trivial.

\end{appendix}
\end{document}